\documentclass[preprint]{aastex6}

\usepackage{amsmath,amssymb}
\usepackage{graphicx}
\usepackage{subfig}
\usepackage{hyperref}

\slugcomment{Submitted for publication in ApJS}
\shorttitle{Relativistic particle integrators}
\shortauthors{Ripperda et al.}

\begin{document}

\title{A comprehensive comparison of relativistic particle integrators}

\author{B. Ripperda\altaffilmark{1}, F. Bacchini\altaffilmark{1}, 
J. Teunissen\altaffilmark{1}, C. Xia\altaffilmark{1}, O. Porth\altaffilmark{2}, L. Sironi\altaffilmark{3}, G. Lapenta\altaffilmark{1}, R. Keppens\altaffilmark{1}}
\altaffiltext{1}{Centre for mathematical Plasma Astrophysics, Department of
Mathematics, KU Leuven, Celestijnenlaan 200B, 3001 Leuven, Belgium}
\altaffiltext{2}{Institut fur Theoretische Physik, Max-von-Laue-Str. 1, D-60438 Frankfurt, Germany}
\altaffiltext{3}{Department of Astronomy, Columbia University, 550 W 120th St, New York, NY 10027, USA}



\begin{abstract}
  We compare relativistic particle integrators commonly used in plasma physics showing several test cases relevant for astrophysics. Three explicit particle pushers are considered, namely the Boris, Vay, and Higuera-Cary schemes. We also present a new relativistic fully implicit particle integrator that is energy conserving. Furthermore, a method based on the relativistic guiding center approximation is included. The algorithms are described such that they can be readily implemented in magnetohydrodynamics codes or Particle-in-Cell codes.
  Our comparison focuses on the strengths and key features of the particle
  integrators. We test the conservation of invariants of motion, and the
  accuracy of particle drift dynamics in highly relativistic, mildly
  relativistic, and non-relativistic settings. The methods are compared in
  idealized test cases, i.e., without considering feedback on the
  electrodynamic fields, collisions, pair creation, or radiation. The test
  cases include uniform electric and magnetic fields, $\mathbf{E}\times\mathbf{B}$-fields, force-free fields, and setups relevant for
  high-energy astrophysics, e.g., a magnetic mirror, a magnetic dipole, and a
  magnetic null. These tests have direct relevance for particle acceleration in
  shocks and in magnetic reconnection.
\end{abstract}

\keywords{acceleration of particles --- relativistic processes --- plasmas --- methods: numerical}

\section{Introduction}
\label{sect:Intro}

In astrophysics, relativistic magnetized flows are ubiquitous around compact objects like black holes and neutron stars. Typical plasma processes in these regimes cover a large range of energy scales, time scales and length scales, from the global fluid scales to the microscopic particle scales. While the large scales can often be captured within the magnetohydrodynamics (MHD) approximation, the small scales cannot. The macroscopic evolution of a plasma often develops relatively slowly, even in relativistic regimes. The macroscopic scale is however tightly coupled to faster phenomena occurring at smaller scales. Many of these phenomena occur in relativistic magnetized plasmas. In the magnetosphere of a compact object, the typical magnetic field can become extremely strong. But in general, even for weaker magnetic fields, the plasma consists of relativistic particles. In these conditions, relativistic effects have to be taken into account for both the global flow and the particles. However, even in the solar corona or the Earth's magnetosphere, where global flows are non-relativistic, particles can accelerate to mildly relativistic energies (\citealt{Li}; \citealt{Ripperda}; \citealt{Ripperda2}).

At relativistic energies the particle equations of motion become nonlinear due to the presence of the Lorentz factor in the momentum of the particle. There are several numerical methods to treat relativistic particle motion accurately. Here we aim to test a selection of available classical and recent explicit leap-frog methods (\citealt{Boris_1970}; \citealt{Vay_2008}; \citealt{HC_2017}). We also present and test a newly implemented fully implicit relativistic method that conserves energy exactly. We apply these methods to known tests for which analytic solutions are available and to more involved setups that are relevant for high-energy astrophysics. Therefore, in the first part of this paper we focus on highly relativistic particles with Lorentz factors much larger than unity, such that the differences between the schemes are well pronounced. We ignore quantum electrodynamics effects, radiation and collisions, and we focus on relativistic particle motion rather than on feedback of the particles to the electromagnetic fields. It is however straightforward to incorporate these effects in MHD or Particle-in-Cell (PIC) codes.

We also compare the obtained particle trajectories to the relativistic guiding center equations of motion. These equations are solved with an explicit fourth-order Runge-Kutta method (\citealt{Ripperda}) and compared to the full solution in order to determine in which regimes gyration can be neglected. The guiding center method has the advantage that the gyration of particles can be neglected, under appropriate assumptions, so that the numerical solution is cheaper to obtain. It also gives additional information on drift motions and acceleration mechanisms for the particles. In the second part of this paper, the comparison of the various integrators, is done in the Newtonian limit, with Lorentz factors close to unity. In this limit all explicit schemes converge to the same solution, and the implicit scheme does too for a sufficiently small time step. The obtained differences in the results can then be unequivocally assigned to the guiding center approximation. 

The accuracy and performance of all methods are tested for various regimes, from Newtonian to highly relativistic energies in idealized setups relevant in astrophysics. Accuracy is assessed by determining how well (approximately) conserved quantities are evolved. This study focuses on the particle pusher and we only consider static, spatially uniform and non-uniform electromagnetic fields. The relativistic pushers considered are commonly used in MHD codes to evolve particles in a global (magnetized) fluid flow (\citealt{Sironi2}; \citealt{Porth3}; \citealt{Ripperda}; \citealt{Ripperda2}) and in Particle-in-Cell codes to evolve both particles and electromagnetic fields (\citealt{Buneman}; \citealt{Spitkovsky}; \citealt{Bowers2008}; \citealt{LapentaMarkidis2011}). In both methods the electromagnetic fields typically have to be interpolated to the particle position. Interpolation errors are tested here, by feeding the pusher with an interpolated, spatially varying field that is known exactly at the particle location, and then we compare to the results for an analytic, spatially varying field.

The particle integrators considered are presented in Section \ref{sect:numerical}. The methods are presented here as independent algorithms and can therefore be readily implemented in any Particle-in-Cell or fluid code to evolve particles interacting with electromagnetic fields. Test cases are presented in Section \ref{sect:testcases_uniform} for uniform fields and in Section \ref{sect:testcases_nonuniform} for non-uniform fields. The guiding center approximation is tested in Section \ref{sect:testcases_gca}. Conclusions are presented in Section \ref{sect:conclusions}.

\section{Numerical methods}
\label{sect:numerical}

In this section we describe the five particle movers used in this paper; the Boris method, the Vay method, the Higuera-Cary method (named HC in the remainder of the paper), the implicit midpoint method and a method based on the guiding center approximation (GCA). We also describe our grid interpolation method. All methods have been implemented to evolve test particles in electromagnetic or magnetohydrodynamic fields obtained from the massively parallel relativistic MHD code {\tt MPI-AMRVAC} (\citealt{Keppensporth}).

All test cases presented here involve charged particles moving in electromagnetic fields $\mathbf{E}$ and $\mathbf{B}$. The relativistic equations of motion for such particles are (in MKS units)
\begin{equation}
\frac{d\mathbf{u}}{dt} = \frac{q}{m} \left(\mathbf{E} + 	\mathbf{v} \times \mathbf{B}\right),
\label{eq:lorentztens}
\end{equation}
and
\begin{equation}
\frac{d\mathbf{x}}{dt}=\mathbf{v},
\label{eq:lorentzpos}
\end{equation}
where $\mathbf{u} = \gamma \mathbf{v}$ is the relativistic momentum vector
divided by the particle rest mass $m$, $\gamma = 1/\sqrt{1-v^2/c^2}$ the Lorentz
factor, $\mathbf{v}$ the velocity, $q$ the charge, and $\mathbf{x}$ the particle position.

Depending on the chosen numerical approach, the equations above are integrated in some discretized form. Below, we present the integration schemes relative to the five methods used in our tests.

\subsection{Explicit leap-frog methods}
\label{sect:explicitleapfrog}

The Boris, Vay, and HC methods are designed to employ a staggered discretization in time for position and velocity of a particle. In essence, the position at some midpoint in time is used to advance the velocity, and the velocity at some staggered point in time drives the motion in space in return. For instance, the velocity can be centered on integer time steps, and the position on half time steps. A discretized version of Equations (\ref{eq:lorentztens})-(\ref{eq:lorentzpos}) reads
\begin{equation}
\frac{\mathbf{u}^{n+1}-\mathbf{u}^{n}}{\Delta t}=\frac{q}{m}\left(\mathbf{E}(\mathbf{x}^{n+1/2})+\bar{\mathbf{v}}\times\mathbf{B}(\mathbf{x}^{n+1/2})\right),
\label{eq:lorentztensdiscstag}
\end{equation}
\begin{equation}
\frac{\mathbf{x}^{n+3/2}-\mathbf{x}^{n+1/2}}{\Delta t}=\mathbf{v}^{n+1},
\label{eq:lorentzposdiscstag}
\end{equation}
where $\bar{\mathbf{v}}$ is some average of the velocity between two timesteps that must be properly defined.
It is often convenient to get rid of the staggering between position and velocity and center both quantities on integer time steps. The scheme remains essentially the same, but the operations can be reordered by splitting the position update in two half steps, one at the end of the current time iteration and the other at the beginning of the next time iteration. This operation is straightforward if one adopts the definition
\begin{equation}
\mathbf{x}^n=\frac{\mathbf{x}^{n+1/2}+\mathbf{x}^{n-1/2}}{2}.
\end{equation}
Such choice results in the sequence of explicit updates
\begin{equation}
\mathbf{x}^{n+1/2} = \mathbf{x}^n + \frac{\mathbf{u}^{n}}{2 \gamma^{n}} \Delta t,
\label{eq:halfposfirst}
\end{equation}
\begin{equation}
\frac{\mathbf{u}^{n+1}-\mathbf{u}^{n}}{\Delta t}=\frac{q}{m}\left(\mathbf{E}(\mathbf{x}^{n+1/2})+\bar{\mathbf{v}}\times\mathbf{B}(\mathbf{x}^{n+1/2})\right),
\label{eq:lorentztensdisc}
\end{equation}
\begin{equation}
\mathbf{x}^{n+1} = \mathbf{x}^{n+1/2} + \frac{\mathbf{u}^{n+1}}{2 \gamma^{n+1}} \Delta t.
\label{eq:halfposlast}
\end{equation}
Note that the average velocity $\bar{\mathbf{v}}$ on the right-hand side of Equation (\ref{eq:lorentztensdisc}) usually involves the unknown $\mathbf{u}^{n+1}$, therefore making the equation implicit. In specific cases, the expression can be formally inverted in order to obtain an explicit expression for $\mathbf{u}^{n+1}$, depending on the choice of $\bar{\mathbf{v}}$. Assuming this is possible, such a modified leap-frog scheme is composed of the following steps:
\begin{itemize}
\item First half of the position update using Equation (\ref{eq:halfposfirst});
\item Explicit solution for $\mathbf{u}^{n+1}$ by analytic inversion of Equation (\ref{eq:lorentztensdisc});
\item Second half of the position update using Equation (\ref{eq:halfposlast}).
\end{itemize}
This ``synchronized'' version of the leap-frog scheme is used for all the tests shown in the next sections. 

The central operation for the solution of the momentum equation is what actually distinguishes each leap-frog method. The Boris, Vay, and HC methods ultimately differ only in the definition of the average velocity, $\bar{\mathbf{v}}$, and therefore in how the analytic inversion is carried out. A fourth explicit second-order method is presented in \cite{Qiang2017_2}. This method is neither time-reversible nor phase-space preserving, but has the aim to perform faster, while maintaining the same accuracy as the Vay integrator. Since we do not consider the computational cost of the schemes here explicitly, we have not considered this integrator in our tests.

All these second-order explicit methods can be extended to fourth-order accuracy, employing the split-operator method (\citealt{Qiang2017}). Such high-order numerical integrators can significantly save the computational cost by using a larger step size as compared with the second-order integrators. The explicit schemes retain their energy-conserving properties at higher orders, since that depends on the formulation of the average velocity, $\bar{\mathbf{v}}$, and not on the order of the scheme (see Appendix \ref{sec:energyconservation}).

The properties of the scheme depend thus strongly on the definition of the average velocity. We are mainly interested in three properties, being 1) energy conservation, 2)  phase space preservation and 3) accurate drift motion.
Energy conservation is discussed in detail in Appendix \ref{sec:energyconservation}, where it is concluded that only one specific choice of average velocity results in strict (numerical) energy conservation. Energy conservation is equivalent to the conservation of the underlying Hamiltonian. Only the implicit scheme presented in Section \ref{subsect:FIalgorithm} conserves the Hamiltonian in the relativistic formulation. The Boris scheme conserves energy in the case of vanishing electric fields. The other schemes do not strictly conserve energy, however in many applications the energy conservation is satisfactory.
Volume preservation is attained by so-called symplectic integrators. Symplectic integrators are designed to conserve symplectic geometries, i.e. areas in phase space: a domain of phase space is mapped by a symplectic function to a new domain of equal area (\citealt{DonnellyRogers2005}). In addition, the energy error is bounded in such methods. Volume preservation is discussed thoroughly in \cite{HC_2017}, concluding that the Boris scheme and the Higuera-Cary scheme are volume preserving but the Vay scheme is not. Volume preservation is defined here as preservation of the differential volume, that is preserved by any solution of the underlying differential equation, with a finite-time-step (\citealt{HC_2017}). We test the preservation of the gyroradius in several different cases of magnetic and electric fields in Section \ref{sect:testcases_uniform}.
The accurate resolution of the drift motion of a particle depends on the problem settings (where \cite{Vay_2008} and \cite{HC_2017} mainly focus on the $\mathbf{E}\times\mathbf{B}$-motion) and is tested thoroughly in Sections \ref{sect:testcases_uniform}, \ref{sect:testcases_nonuniform} and \ref{sect:testcases_gca}.


\subsubsection{Boris method}
\label{subsect:Boris}


The Boris method (\citealt{Boris_1970}) is a classic, second order accurate leap-frog
scheme that is widely used. Even though it was first described almost 50 years
ago, the method is still actively investigated
(\citealt{Vay_2008,Qin_2013,Ellison_2015}), in particular its volume-preserving and
symplectic properties. 
For the Boris method, the definition of the average velocity is
\begin{equation}
\bar{\mathbf{v}} = \frac{\mathbf{u}^{n+1} + \mathbf{u}^n}{2 \gamma^{n+1/2}}.
\label{eq:vbarBoris}
\end{equation}
The inversion step is then given by the following operations (see e.g. \citealt{Birdsall}):
\begin{itemize}
\item First half electric field acceleration:
\begin{equation}
  \mathbf{u}^{-} = \mathbf{u}^{n} + \frac{q\Delta t}{2 m}\mathbf{E}(\mathbf{x}^{n+1/2}).
\end{equation}
\item Rotation step:
\begin{equation}
\mathbf{u}^{+} = \mathbf{u}^{-} + (\mathbf{u}^{-} + (\mathbf{u}^{-} \times \mathbf{t})) \times \mathbf{s}.
\end{equation}
\item Second half electric field acceleration:
\begin{equation}
\mathbf{u}^{n+1} = \mathbf{u}^{+} + \frac{q\Delta t}{2 m}\mathbf{E}(\mathbf{x}^{n+1/2}).
\end{equation}
\end{itemize}
Here, the auxiliary quantities are $\gamma^{-} = \sqrt{1+(u^{-}/c)^2}$, $\gamma^{+} = \sqrt{1+(u^{+}/c)^2} = \gamma^{-}$, $\boldsymbol{t} = \mathbf{B}(\mathbf{x}^{n+1/2})q\Delta t/(2 m\gamma^{-})$, $ \mathbf{s} = 2\boldsymbol{t}/(1+t^2)$. The pure rotation of the velocity vector to obtain $\mathbf{u}^{+}$ results in the Lorentz factor at the midstep $\gamma^{n+1/2} = \gamma^{-} = \gamma{^+}$. In Appendix \ref{sec:energyconservation} this property is used to show that the Boris scheme is energy conserving in case of pure magnetic fields, meaning that it preserves the property that magnetic fields do not exert work.  By setting $\gamma \equiv 1$, the scheme somewhat simplifies and one obtains a non-relativistic (Newtonian) version. 
When the magnetic field strength varies in space, it becomes attractive to use
an adaptive time step. The time-symmetry of the scheme is then lost (see e.g.
\citealt{Hairer_1997}). Our implementation in \texttt{MPI-AMRVAC} supports
adaptive time stepping, which has been implemented with the synchronized version
of the scheme, since then the same $\Delta t$ can be used for both the position
and velocity update. However, for the tests presented here we have employed a
fixed time step.


\subsubsection{Vay method}
\label{subsect:Vay}

To counteract spurious acceleration of the particles by perpendicular electric fields \cite{Vay_2008} proposed a modification of the Boris algorithm by defining the average velocity as
\begin{equation}
\bar{\mathbf{v}} = \frac{\mathbf{u}^{n}/\gamma^n + \mathbf{u}^{n+1}/\gamma^{n+1} }{2}.
\label{eq:Vayaveragev}
\end{equation}
The analytic inversion of equation (\ref{eq:lorentztensdisc}) is done in two steps:
\begin{itemize}
\item Field contribution:
\begin{equation}
\mathbf{u}^{n+1/2} = \mathbf{u}^n + \frac{q \Delta t}{2 m}\left(\mathbf{E}(\mathbf{x}^{n+1/2}) + \frac{\mathbf{u}^n}{\gamma^n} \times \mathbf{B}(\mathbf{x}^{n+1/2})\right).
\end{equation}
\item Rotation step:
\begin{equation}
\mathbf{u}^{n+1} = s\left[\mathbf{u}' + \left(\mathbf{u}' \cdot \mathbf{t} \right) \mathbf{t} + \mathbf{u}' \times \mathbf{t}\right].
\end{equation}
\end{itemize}
Here, the auxiliary quantities are given by $\mathbf{u}' = \mathbf{u}^{n+1/2} +  \mathbf{E}(\mathbf{x}^{n+1/2}) q \Delta t / (2m)$, $\boldsymbol{\tau} = \mathbf{B}(\mathbf{x}^{n+1/2}) q \Delta t/(2 m)$, $u^{\ast} = \mathbf{u}' \cdot \boldsymbol{\tau} / c$, $\sigma = \gamma'^{2} - \tau^2$, $\gamma' = \sqrt{1 + u'^2/c^2}$, $\mathbf{t} = \boldsymbol{\tau} / \gamma^{n+1}$, and $s = 1/(1+t^2)$, with
\begin{equation}
\gamma^{n+1} = \sqrt{\frac{\sigma + \sqrt{\sigma^2 + 4\left(\tau^2 + (u^{\ast})^2\right)}}{2}}.
\label{eq:gammaVay}
\end{equation}
The position update is done in accordance with the implemented Boris scheme by performing half the position update at the end of a step and the other half at the beginning of the next step.
\subsubsection{Higuera-Cary method}
The Boris scheme is known to be volume-preserving, meaning that gyration is accurately resolved. The Vay scheme is an adaptation of the Boris scheme designed to preserve the $\mathbf{E}\times\mathbf{B}$-velocity, which is not correctly computed with the Boris method. The Vay method is not volume-preserving, which can lead to a larger error in the gyroradius. \cite{HC_2017} proposed a new volume-preserving method that also resolves the $\mathbf{E}\times\mathbf{B}$ motion accurately, while keeping the computational cost similar to the Vay scheme. The method is claimed to conserve energy and is shown to resolve typical idealized astrophysical test cases with more accuracy than the Boris and Vay methods (\citealt{HC_2017}). The scheme relies on a new choice of the average velocity as
\begin{equation}
\bar{\mathbf{v}} = \frac{\mathbf{u}^{n+1} + \mathbf{u}^n}{2\bar{\gamma}},
\label{eq:vbarhc}
\end{equation}
with
\begin{equation}
\bar{\gamma} = \sqrt{1 +\left(\frac{\mathbf{u}^{n+1} + \mathbf{u}^n}{2c}\right)^2}.
\label{eq:gammahc}
\end{equation}
For this choice of $\bar{\mathbf{v}}$, the analytic inversion of Equation (\ref{eq:lorentztensdisc}) is performed in three steps:
\begin{itemize}
\item First half electric field acceleration:
\begin{equation}
\mathbf{u}^{-} = \mathbf{u}^{n} + \frac{q\Delta t}{2 m}\mathbf{E}(\mathbf{x}^{n+1/2}).
\end{equation}
\item Rotation step:
\begin{equation}
\mathbf{u}^{+} = s\left[\mathbf{u}^- + \left(\mathbf{u}^- \cdot \mathbf{t} \right) \mathbf{t} + \mathbf{u}^- \times \mathbf{t}\right].
\end{equation}
\item Second half electric field acceleration:
\begin{equation}
\mathbf{u}^{n+1} = \mathbf{u}^{+} + \frac{q\Delta t}{2 m}\mathbf{E}(\mathbf{x}^{n+1/2})+\mathbf{u}^-\times\mathbf{t}.
\end{equation}
\end{itemize}
Here, the auxiliary quantities are $\gamma^{-} = \sqrt{1+(\mathbf{u}^{-})^2/c^2}$, $\boldsymbol{\tau} = \mathbf{B}(\mathbf{x}^{n+1/2}) q \Delta t  / (2m)$, $u^{\ast} = \mathbf{u}^{-} \cdot \boldsymbol{\tau}/c$, $\sigma = (\gamma^{-})^2 - \tau^2$, $\mathbf{t} = \boldsymbol{\tau}/\gamma^{+}$, and $s = 1/(1+t^2)$, with
\begin{equation}
\gamma^{+} = \sqrt{\frac{\sigma+\sqrt{\sigma^2+4(\tau^2+(u^{\ast})^2)}}{2}}.
\end{equation}
Once again, the position update is done in accordance with the implemented Boris scheme by performing half the position update at the end of a
step and the other half at the beginning of the next step.


\subsection{Implicit midpoint method}
\label{subsect:FIalgorithm}

The explicit schemes presented above allow for fast solution of the particle
motion. The average velocity, $\bar{\mathbf{v}}$, is chosen such that
Equation (\ref{eq:lorentztensdisc}) can be analytically inverted to retrieve an
explicit expression for $\mathbf{u}^{n+1}$. However, the split form of the
position update (in our synchronized leap-frog formulation) shows that there is
a discrepancy between the average velocity used to advance $\mathbf{u}$, which is $\bar{\mathbf{v}}$ and that
used for $\mathbf{x}$. By combining Equations (\ref{eq:halfposfirst}) and
(\ref{eq:halfposlast}) it follows that
\begin{equation}
\frac{\mathbf{x}^{n+1}-\mathbf{x}^n}{\Delta t}=\frac{1}{2}\left(\frac{\mathbf{u}^{n+1}}{\gamma^{n+1}}+\frac{\mathbf{u}^n}{\gamma^n}\right).
\end{equation}
Thus the update of the position is, in general, driven by an average velocity that may differ from the chosen form used in Equation (\ref{eq:lorentztensdisc}). 
A consistent form of the system of equations addressing the latter issue reads
\begin{equation}
\frac{\mathbf{u}^{n+1}-\mathbf{u}^{n}}{\Delta t}=\frac{q}{m}\left(\mathbf{E}(\mathbf{x}^{n+1/2})+\bar{\mathbf{v}}\times\mathbf{B}(\mathbf{x}^{n+1/2})\right),
\label{eq:lorentztensdiscimpl}
\end{equation}
\begin{equation}
\frac{\mathbf{x}^{n+1}-\mathbf{x}^{n}}{\Delta t}=\bar{\mathbf{v}},
\label{eq:lorentzposdiscimpl}
\end{equation}
where the update of the position is now driven by the same average velocity used in the momentum equation. The two equations are coupled through 
\begin{equation}
 \mathbf{x}^{n+1/2}=\frac{\mathbf{x}^{n+1}+\mathbf{x}^n}{2},
\end{equation}
and the resulting system of nonlinear equations can be reduced to a 3-dimensional system by the relation
\begin{equation}
 \mathbf{x}^{n+1/2}=\mathbf{x}^n+\frac{\Delta t}{2}\bar{\mathbf{v}},
 \label{eq:xbarN}
\end{equation}
which differs from equation \ref{eq:halfposfirst}. For an arbitrary choice of $\bar{\mathbf{v}}$, it is not possible to formally invert Equation (\ref{eq:lorentztensdiscimpl} )and therefore one has to solve for $\mathbf{u}^{n+1}$ with an iterative method.
 
In our tests, we adopt the expression for the average velocity proposed by \cite{LapentaMarkidis2011},
\begin{equation}
 \bar{\mathbf{v}}=\frac{\mathbf{u}^{n+1}+\mathbf{u}^n}{\gamma^{n+1}+\gamma^n},
 \label{eq:vbarN}
\end{equation}
which differs from the expressions used in the Boris, Vay, and Higuera-Cary schemes. An interesting property that arises with the above definition is the conservation of energy to machine precision (\citealt{LapentaMarkidis2011}). This property suppresses spurious particle heating in, e.g. Particle-in-Cell simulations, avoiding typical  numerical instabilities characterizing explicit schemes (\citealt{Birdsall}). This is not ensured in the case of other choices for the average velocity (see Appendix \ref{sec:energyconservation} for a formal proof of the energy conservation, or a lack thereof, of all schemes). \cite{Petri2017} presents a fully implicit scheme similar to the one above, but with a different choice of average velocity (specifically, the one used in \citealt{Vay_2008}).
The solution of system (\ref{eq:lorentztensdiscimpl}) can be carried out with several approaches (see e.g. \citealt{Noguchi2007}). In this work we choose to adopt a Newton algorithm similar to the one presented in \cite{Siddi}, but extended here to the relativistic case. The solution algorithm is composed, at each time iteration, of the following steps:
\begin{itemize}
 \item The nonlinear cycle is initialized by assigning to the unknown, $\mathbf{u}_k$, an initial guess $\mathbf{u}^*$, where $k$ over the iterations. In our implementation we find that using the value at the previous time step, $\mathbf{u}^n$ proves satisfactory for all tests.
 \item The $k$-th average velocity, $\bar{\mathbf{v}}_k$, and position, $\bar{\mathbf{x}}_k = \mathbf{x}_k^{n+1/2}$, are computed using $\mathbf{u}_k$ as the value for $\mathbf{u}^{n+1}$. The field values, $\mathbf{E}$ and $\mathbf{B}$, are interpolated at $\bar{\mathbf{x}}_k$ .
 \item The $k$-th residual is computed according to
 \begin{equation}
   \mathbf{F}(\mathbf{u}_k)=\mathbf{u}_k-\mathbf{u}^n-\frac{q\Delta t}{m}\left(\mathbf{E}(\bar{\mathbf{x}}_k)+\frac{\mathbf{u}_k+\mathbf{u}^n}{\gamma_k+\gamma^n}\times\mathbf{B}(\bar{\mathbf{x}}_k)\right),
   \label{eq:residual}
 \end{equation}
 where $\gamma_k=\sqrt{1+u_k^2/c^2}$. The Jacobian is obtained accordingly as $J(\mathbf{u}_k)=\partial\mathbf{F}(\mathbf{u}_k)/\partial\mathbf{u}_k$ by analytic differentiation of Equation \ref{eq:residual} above.
 \item The iteration variable is updated by solving the linear system, $J(\mathbf{u}_k)(\mathbf{u}_{k+1}-\mathbf{u}_k)=-\mathbf{F}(\mathbf{u}_k)$.
 \item A termination criterion is applied, e.g. by evaluating $|\mathbf{u}_{k+1}-\mathbf{u}_k|<\textrm{tol}$, where $\textrm{tol}$ is some tolerance value. We choose $\textrm{tol}=10^{-14}$, slightly above double precision round off, for our tests. If the stop criterion is met, the cycle is terminated, otherwise it is restarted by assigning $\mathbf{u}_k=\mathbf{u}_{k+1}$.
\end{itemize}
When the stop criterion is met, the new proper velocity is taken as $\mathbf{u}^{n+1}=\mathbf{u}_{k+1}$. Finally, the new particle position is updated according to
\begin{equation}
\mathbf{x}^{n+1}=\mathbf{x}^n+\frac{\mathbf{u}^{n+1}+\mathbf{u}^n}{\gamma^{n+1}+\gamma^n}\Delta t.
\end{equation}
In computing the Jacobian, $J(\mathbf{u}_k)$, it is necessary to evaluate the derivatives of the field terms in Equation (\ref{eq:residual}). In the most general case where such terms are obtained via interpolation from a grid, the derivatives of the fields reduce to derivatives of the chosen interpolation function. In this work we choose to evaluate the fields via linear interpolation, hence dedicated routines are used to compute the corresponding interpolated derivative of the fields at the particle position. Given the residual functions (\ref{eq:residual}), each term in the Jacobian matrix is usually a fairly complicated expression. For simplicity, but without loss of generality, consider the $x-$component of Equation \ref{eq:residual}, which reads
\begin{equation}
 F_x(\mathbf{u})=u_x-u_x^n-\frac{q\Delta t}{m}\left(E_x(\bar{\mathbf{x}})+\frac{u_y+u_y^n}{\gamma+\gamma^n} B_z(\bar{\mathbf{x}})-\frac{u_z+u_z^n}{\gamma+\gamma^n} B_y(\bar{\mathbf{x}})\right).
\end{equation}
Then the first element of the Jacobian, $J_{xx}(\mathbf{u})=\partial F_x(\mathbf{u})/\partial u_x$, is given by
\begin{equation} 
 \begin{split}
 J_{xx}(\mathbf{u})=1-\frac{q\Delta t}{m} \biggl[ & \frac{\partial {E}_x(\bar{\mathbf{x}})}{\partial u_{x}}+\frac{u_y+u_y^n}{\gamma+\gamma^n}\frac{\partial {B}_z(\bar{\mathbf{x}})}{\partial u_{x}}-\frac{u_z+u_z^n}{\gamma+\gamma^n}\frac{\partial {B}_y(\bar{\mathbf{x}})}{\partial u_{x}} \\
  & +C_{2,x}(u_{y}+u^n_y){B}_z(\bar{\mathbf{x}})-C_{2,x}(u_{z}+u^n_z){B}_y(\bar{\mathbf{x}}) \biggr],
 \end{split}
 \label{eq:J11}
\end{equation}
where the derivatives of the field terms are computed via the chain rule, such that, e.g. for the electric field,
\begin{equation}
 \frac{\partial {E}_x(\bar{\mathbf{x}})}{\partial u_{x}}=
 \frac{\partial {E}_x(\bar{\mathbf{x}})}{\partial\bar{x}}\frac{\partial \bar{x}}{\partial u_{x}} +
 \frac{\partial {E}_x(\bar{\mathbf{x}})}{\partial\bar{y}}\frac{\partial \bar{y}}{\partial u_{x}} +
 \frac{\partial {E}_x(\bar{\mathbf{x}})}{\partial\bar{z}}\frac{\partial \bar{z}}{\partial u_{x}}.
\end{equation}
This expression is convenient, since the fields are functions of the position, allowing one to use analytic derivatives (if the fields are given by an analytic expression) or take the derivative of the interpolation functions (if the fields are retrieved via interpolation). From Equations \ref{eq:xbarN} and \ref{eq:vbarN}, the derivatives of mid-point coordinates with respect to the dimensionless 4-velocity further reduce to
\begin{equation}
 \frac{\partial \bar{x}}{\partial u_{x}}=\frac{\Delta t}{2}C_{1,x},
\end{equation}
\begin{equation}
 \frac{\partial \bar{y}}{\partial u_{x}}=\frac{\partial \bar{z}}{\partial u_{x}}=\frac{\Delta t}{2}C_{2,x},
\end{equation}
where
\begin{equation}
 C_{1,i}=\frac{\partial}{\partial u_{i}}\left(\frac{u_{i}+u_i^n}{\gamma+\gamma^n}\right)=\frac{\gamma+\gamma^n-u_{i}(u_{i}+u_i^n)/(\gamma c^2)}{(\gamma+\gamma^n)^2},
\end{equation}
\begin{equation}
 C_{2,i}=\frac{\partial}{\partial u_{i}}\left({\gamma+\gamma^n}\right)^{-1}=-\frac{u_{i}/(\gamma c^2)}{(\gamma+\gamma^n)^2},
\end{equation}
which also define the coefficients in Equation \ref{eq:J11} above. The exact same reasoning leads to the expressions of the other terms in the Jacobian matrix.

It is well known that the Newton algorithm is not guaranteed to converge. In some pathological cases, e.g. when the Jacobian vanishes at the solution, the iteration fails to converge; other non-convergence issues arise for bad choices of the initial guess or fast oscillations of the residual function around the solution. In practice, however, non-convergence is very rarely observed for calculations in which a relatively small time step ensures that quantities do not change abruptly from one time level to the next. Nevertheless, this should not be regarded as an absolute limitation on the choice of $\Delta t$. For systems evolving according to fast dynamics, the user will be interested in capturing the details of the evolution, thus preferring a small time step; conversely, for slowly-evolving systems, the algorithm is likely to converge since the change in the variables from one time level to the next will not be abrupt. A suitable choice of the initial guess represents a crucial factor in ensuring the convergence of the algorithm, and should be considered carefully. When convergence is reached, in most cases the convergence rate is second order (\citealt{Press}). In our tests, we typically observe convergence to the chosen $10^{-14}$ absolute tolerance within 4-5 iterations, when using the values of $\mathbf{u}$ at the previous time step as an initial guess for the Newton step. It is important to note that, instead of using the classical Newton algorithm with analytical Jacobian, Jacobian-free methods (e.g. Newton-Krylov solvers, see \citealt{SaadSchultz1986}) could be adopted. The advantage of not having to compute the Jacobian, however, comes at the cost of a typically higher number of iterations needed to reach convergence.

The implicit method has the important property that it avoids the decoupling of the magnetic field advance and the electric field advance that is typical for the explicit methods. As shown by \cite{Vay_2008}, this decoupling leads to a break of Lorentz invariance and the introduction of spurious forces, for the Boris method. The Vay method and the HC method are proven to maintain their Lorentz invariance (\citealt{Vay_2008}; \citealt{HC_2017}). The implicit algorithm does not decouple the electric and magnetic field advance, avoiding the problem of maintaining Lorentz invariance completely, as demonstrated in \cite{LapentaMarkidis2011}.


\subsection{The guiding center approximation}
\label{subsect:GCA}

In certain astrophysical circumstances the typical length scale of the gradient in the magnetic field $L$ is large compared to the gyroradius $R_c$ of the particle. In this case the gyration can be neglected for test particles. The center of the gyration (or, guiding center) is evolved rather than the actual particle position and the equations of motion simplify significantly, allowing for a less expensive numerical solution. The guiding center approximation is applied to Equation (\ref{eq:lorentztens}) to obtain the relativistic guiding center equations of motion describing the (change in) guiding center position $\mathbf{R}$, parallel relativistic momentum $m\gamma v_{\|}$ and relativistic magnetic moment $\mu_r = m \gamma^{2} v^{2}_{\perp}/2B$ in three-space (\citealt{Vandervoort})
\begin{align*}
\frac{d\mathbf{R}}{dt} = v_{\|}\hat{\mathbf{b}} - \frac{\hat{\mathbf{b}} \times c \mathbf{E}}{B} + 
\end{align*}
\begin{align*}
\frac{\hat{\mathbf{b}}}{B\left(1-\frac{E_{\perp}^{2}}{B^2}\right)} \times \Biggl\{\frac{cm\gamma}{q}\left(v_{\|}^{2}\left(\hat{\mathbf{b}}\cdot\nabla\right)\hat{\mathbf{b}}+v_{\|}\left(\mathbf{u}_E\cdot\nabla\right)\hat{\mathbf{b}} + v_{\|}\left(\hat{\mathbf{b}}\cdot\nabla\right)\mathbf{u}_E + \left(\mathbf{u}_E\cdot \nabla\right)\mathbf{u}_E\right) + \nonumber
\end{align*}
\begin{align}
\Biggl. \frac{\mu_r c}{\gamma q}\nabla\left[B\left(1-\frac{E_{\perp}^{2}}{B^2}\right)^{1/2}\right]  + \frac{v_{\|}E_{\|}}{c}\mathbf{u}_E  \Biggr\},
\label{eq:gcastatic1}
\end{align}
\begin{align}
\frac{d \left(m \gamma v_{\|}\right)}{dt} =  m\gamma\mathbf{u}_E\cdot \left(v_{\|}\left(\hat{\mathbf{b}}\cdot\nabla\right)\hat{\mathbf{b}}+\left(\mathbf{u}_E\cdot\nabla\right)\hat{\mathbf{b}}\right) +\nonumber \\ 
qE_{\|} -\frac{\mu_r}{\gamma}\hat{\mathbf{b}}\cdot\nabla\left[B\left(1-\frac{E^{2}_{\perp}}{B^2}\right)^{1/2}\right],
\label{eq:gcastatic2}
\end{align}
\begin{equation}
\frac{d \left(m \gamma^{*2} v^{*2}_{\perp}/2B^*\right)}{dt} = \frac{d \mu_{r}^{*}}{dt} = 0.
\label{eq:gcastatic3}
\end{equation}
Here, $\hat{\mathbf{b}}$ is the unit vector in the direction of the magnetic field and $v_{\|}$ the component of the particle velocity vector parallel to $\hat{\mathbf{b}}$. The magnitude of the electric field is split as $E = \sqrt{E^{2}_{\perp} +E^{2}_{\|}}$, with $E_{\perp}$ the component of the electric field perpendicular to $\mathbf{B}$ and $E_{\|}$ the parallel component. The drift velocity, perpendicular to $\mathbf{B}$ is written as $\mathbf{u}_E = c\mathbf{E}\times\hat{\mathbf{b}}/B$ and $v^{*}_{\perp}$ is the perpendicular velocity of the particle, in the frame of reference moving at $\mathbf{u}_E$. The magnetic field in that frame is given by $B^* = B(1-E^{2}_{\perp}/B^2)^{1/2}$ up to first order. The relativistic magnetic moment $\mu_{r}^{*}$ is an adiabatic invariant and is proportional to the magnetic flux through the gyration circle, again in the frame of reference moving at $\mathbf{u}_E$. The oscillation of the Lorentz factor at the gyrofrequency is averaged out as well, giving $\gamma = \gamma^{*}(1-E^{2}_{\perp}/B^2)^{-1/2}$. We assume the electromagnetic fields to be slowly varying compared to the particle dynamics, such that no temporal derivatives of electromagnetic fields appear in Equations (\ref{eq:gcastatic1})-(\ref{eq:gcastatic3}).

\subsubsection{Particle drifts}
\label{sect:newtonian}

For the guiding center method we store all the field-dependent terms in Equations (\ref{eq:gcastatic1}) and (\ref{eq:gcastatic2}) as grid variables, and then linearly interpolate them at the location of the guiding center. This was done to improve efficiency and to avoid having to use a wider interpolation stencil to determine gradients. Having the terms in the GCA equations available provides the opportunity to obtain information about particle drifts. Every term in Equation (\ref{eq:gcastatic1}) represents a drifting motion of the particle and every term in (\ref{eq:gcastatic2}) represents an acceleration mechanism. The meaning of these terms becomes clearer in the Newtonian approximation where $v^2 \ll c^2$ and the magnitude of its $\mathbf{E}\times\mathbf{B}$-drift velocity $u_E^2 \ll c^2$ such that $\gamma \rightarrow 1$ and $1/\sqrt{(1-E_{\perp}^2/B^2)} = 1/\sqrt{1-u_{E}^2/c^2} \rightarrow 1$. Then also the relativistic magnetic moment, a constant of motion, becomes the classical magnetic moment $\mu_r = m\gamma^2 v_{\perp}^2/2B \rightarrow \mu = m v_{\perp}^2/2B$. Applying this limit gives the Newtonian equations of motion for the guiding center
\begin{equation}
\frac{d\mathbf{R}}{dt} =v_{\|}\hat{\mathbf{b}}+\frac{\hat{\mathbf{b}}}{B} \times \Biggl\{ -c\mathbf{E} + \frac{cm}{q}\left(v_{\|}\frac{d\hat{\mathbf{b}}}{dt}+\frac{d\mathbf{u}_E}{dt}\right)  + \frac{\mu c}{q}\nabla B \Biggr\},
\label{eq:gcanewton1}
\end{equation}
\begin{equation}
\frac{d \left(m v_{\|}\right)}{dt} = m \mathbf{u}_E\cdot \frac{d\hat{\mathbf{b}}}{dt} +qE_{\|} -\mu\hat{\mathbf{b}}\cdot\nabla B.
\label{eq:gcanewton2}
\end{equation}
The first term on the right-hand-side of Equations (\ref{eq:gcastatic1}) and (\ref{eq:gcanewton1}) is the motion parallel to $\hat{\mathbf{b}}$ following from the solution of Equations (\ref{eq:gcastatic2}) and (\ref{eq:gcanewton2}) respectively. The second term in Equations (\ref{eq:gcastatic1}) and (\ref{eq:gcanewton1}) is the $\mathbf{E}\times\mathbf{B}$ drift. The third term combines the curvature drift (resulting from the static part of the inertial drift) $v_{\|}d\hat{\mathbf{b}}/dt = v_{\|}^{2}\left(\hat{\mathbf{b}}\cdot\nabla\right)\hat{\mathbf{b}}+v_{\|}\left(\mathbf{u}_E\cdot\nabla\right)\hat{\mathbf{b}}$ and the polarisation drift $d\mathbf{u}_E/dt = v_{\|}\left(\hat{\mathbf{b}}\cdot\nabla\right)\mathbf{u}_E + \left(\mathbf{u}_E\cdot \nabla\right)\mathbf{u}_E$, where non-static fields are neglected. For these drifts the gyration period increases by a factor $\gamma$ in the relativistic Equation (\ref{eq:gcastatic1}), resulting from the effective mass of the gyrating particle $\gamma m$. Then, the magnitude of the magnetic field, in the frame of reference moving at $\mathbf{u}_E$, is $B^* = B(1-E^{2}_{\perp}/B^2)^{1/2} = B\sqrt{1-u_{E}^2/c^2}$ up to first order, explaining the factor $1/\sqrt{1-u_{E}^2/c^2}$ appearing in all terms including the magnetic field magnitude in relativistic Equations  (\ref{eq:gcastatic1}) and (\ref{eq:gcastatic2}). The fourth term in Equations (\ref{eq:gcastatic1}) and (\ref{eq:gcanewton1}) is the $\nabla B$ drift, also a factor $\gamma$ larger than in Equation (\ref{eq:gcanewton1}). The last term on the right-hand-side of Equation (\ref{eq:gcastatic1}) is an additional, purely relativistic drift in the direction $\hat{\mathbf{b}} \times \mathbf{u}_E$ that is negligible in the Newtonian limit and does not appear in Equation (\ref{eq:gcanewton1}) (\citealt{Northropbook}). The differences between Newtonian and relativistic guiding center equations of motion are discussed in more detail in \cite{Ripperda}.

\subsubsection{Runge-Kutta method}
Equations (\ref{eq:gcastatic1}-\ref{eq:gcastatic3}) are advanced with a fourth order Runge-Kutta scheme with adaptive time stepping. Here, the particle timestep $\delta t$ is determined based on its parallel acceleration $a = d v_{\|}/dt$ and velocity $v = \sqrt{(v_{\|})^2 + (v_{\perp})^2}$ as the minimum of $\delta r / v$ and $v / a$, where $\delta r$ is the grid step. This time step is restricted such that a particle cannot cross more than one grid cell in one time step. The fields $\mathbf{E}$ and $\mathbf{B}$, and for the GCA equations their spatial derivatives, are obtained at the particle position via linear interpolations in space or they are given analytically. The particle gyroradius is also calculated at every timestep and compared to the typical cell size to monitor the validity of the guiding center approximation.




\subsection{Grid interpolation}

The particle movers in {\tt MPI-AMRVAC} can get the electric and magnetic field
at a location in two ways: from a user-defined routine (e.g. an analytic
function), or by (bi/tri)linear interpolation from grid variables. For the test
cases presented here involving static fields with linear spatial gradients, both
methods yield the same results. For more general fields there will be an
interpolation error proportional to $\Delta x^2$, i.e., the square of the grid
spacing. We remark that for time-varying fields, which are not
  considered here, {\tt MPI-AMRVAC} also performs linear interpolation in time.

For smooth fields, the use of higher-order spatial and temporal
  interpolation methods can greatly reduce interpolation errors. This would be
  particularly attractive when high-order reconstruction methods are used to
  compute electric and magnetic fields (see e.g.
  \citealt{Balsara_2009,Balsara_2017}). The reconstruction methods can then be
  re-used to interpolate the solution. However, a higher-order interpolation
  method requires a wider numerical stencil, which complicates the
  implementation near grid boundaries (\citealt{Borovikov_2015}). Furthermore, a
  limiting procedure is required to avoid unrealistic interpolation values near
  large gradients.

{\tt MPI-AMRVAC} divides the computational grid in blocks, which are distributed over
the processors. Each block also contains a few layers of \emph{ghost cells},
with data from neighboring blocks. In our current implementation the time step
is restricted so that particles cannot move more than one cell out of their
current grid block. After every step of the particle movers, particles are
moved to a different processor if required.



\subsection{Computational cost of schemes}
\label{sec:computational-cost}

The implementation of the schemes described above has not yet been optimized for
performance. Nevertheless, we here try to give a rough idea of the relative cost
of the different methods. There are a number of factors to consider when judging
the computational cost of a scheme, which are discussed below.

\paragraph{Time step restrictions}
The main advantage of the guiding center approximation is that its time step is
not limited by the gyration time, since gyration is neglected. All the other schemes require a time step that
is a fraction of the gyration time.

\paragraph{Cost of numerically solving the scheme's equations} The Boris method
and its derivatives (Vay, Higuera-Cary) have about the same cost. For the
guiding center approximation, the equations have more terms and the cost is
higher, but the procedure is still explicit. For the implicit method, the cost
is about 4--5 times that of the Boris method. For reference, we have measured
the performance of the schemes for the gyration test described in section
\ref{subsect:gyration}. Advancing $10^3$ particles over $10^4$ time steps on a
single $2.4 \, \textrm{GHz}$ processor took about $50 \, \textrm{s}$ with the
Boris, Vay, and Higuera-Cary methods, and $2.2 \times 10^2 \, \textrm{s}$ with the fully-implicit method. 
\paragraph{Interpolation costs} In order to obtain the fields at a particle
position, data has to be loaded from memory and interpolated. We here use linear
interpolation, for which the numerical computations are relatively cheap
compared to the cost of loading data from memory. Optimizations to improve data
locality, for example by sorting particles based on their position in a grid block, have yet to be made in {\tt MPI-AMRVAC}.

\paragraph{Parallelization} The tests presented in this paper were all performed
on a single processor, although our implementation also allows for parallel
runs. However, we currently use the domain decomposition used for the MHD
simulations to distribute the particles (i.e., load balancing does not account for test particles), which can lead to unequal load
balancing and requires frequent synchronization.



\section{Test cases}
In this section we test all the schemes presented above in idealized setups that are relevant for astrophysics. In all tests we employ a uniform, rectilinear 3D Cartesian grid to store the field values. The grid resolution is coarse ($16\times 16\times 16$) in the case of uniform fields, since interpolation does not affect the results. For nonuniform fields, when interpolation is used, the resolution is specified in the corresponding sections. We adopt SI units, which are omitted in the text and figures, such that electric field $\mathbf{E}$ is in $[V m^{-1}]$, magnetic field $\mathbf{B}$ in $[T]$, particle charge $q$ in $[C]$ and mass $m$ in $[kg]$. The particle velocity $\mathbf{v}$ is in $[m s^{-1}]$ and the position $\mathbf{x}$ in $[m]$.

We first consider five relativistic test cases, as summarized in
  Table \ref{tab:example_table}. Then we investigate the error in the guiding
  center approximation by comparing it to Boris method in three test cases
  ($\nabla B$-drift, magnetic null and magnetic dipole), all in the
  non-relativistic regime, i.e., $\gamma \rightarrow 1$.

\begin{table}
	\centering
	\caption{Overview of relativistic test cases. The symbols indicate that
          a method has a small error (compared to the other schemes) in the
          Lorentz factor ($\gamma$), the velocity ($\mathbf{v}$), the gyroradius
          ($R_c$), or the gyro-phase ($\theta_c$).}
	\label{tab:example_table}
	\begin{tabular}{lcccr} 
		\hline
		test & Boris & Vay & HC & Implicit \\
		\hline
		uniform $\mathbf{E}$-field & $\gamma$ & $\gamma$ & $\gamma$ & $\gamma$  \\
		uniform $\mathbf{B}$-field & $\gamma$ &  & $\theta_c$ & $\gamma$ \\
		Force-free field & $\gamma$& $\gamma$, $\mathbf{v}$ & $\gamma$ & $\gamma$, $\mathbf{v}$ \\
		$\mathbf{E} \times \mathbf{B}$-drift &  &  &  &$\gamma$, $R_c$ \\
		magnetic mirror & $\gamma$ &  &  & $\gamma$ \\

		\hline
	\end{tabular}
\end{table}

\subsection{Uniform static fields}
\label{sect:testcases_uniform}

\subsubsection{Uniform electric field}

A charged particle in a uniform electric field only experiences acceleration in the direction of $\mathbf{E}$. For a constant $\mathbf{E}=(E_x,0,0)$, the relativistic equations of motion (\ref{eq:lorentztens})-(\ref{eq:lorentzpos}) can be solved analytically. For a particle initially at rest at the origin of the coordinate system, the result is
\begin{equation}
 x_\mathrm{an}(t)=\frac{m c^2}{q E}(\gamma_\mathrm{an}(t)-1),
\end{equation}
\begin{equation}
v_{x,\mathrm{an}} (t)=\frac{q E}{m} \frac{t}{\gamma_\mathrm{an}(t)}, 
\end{equation}
where $\gamma_\mathrm{an}(t)=\sqrt{1+(qEt)^2/(mc)^2}$. At late times $t\gg mc/(|q|E)$, the growth of the Lorentz factor is nearly linear, whereas the velocity $v_{x,\mathrm{an}} \approx c$.

We use the same setup described in \cite{Petri2017} in order to simulate the extreme acceleration of a particle with charge $q=1$ and mass $m=1$, up to a Lorentz factor of order $\sim 10^9$. For this purpose, we set up a uniform electric field $E_x/c=1$, with a particle initially at rest at $\mathbf{x}(t=0)=(0,0,0)$. We let the simulation run up to $t=10^9$ with a time step $\Delta t=10^3$. The experiment is repeated for each integration method. The  results can be directly compared to the analytic solutions above.

\begin{figure}[!h]
\centering
\includegraphics[width=0.5\columnwidth]{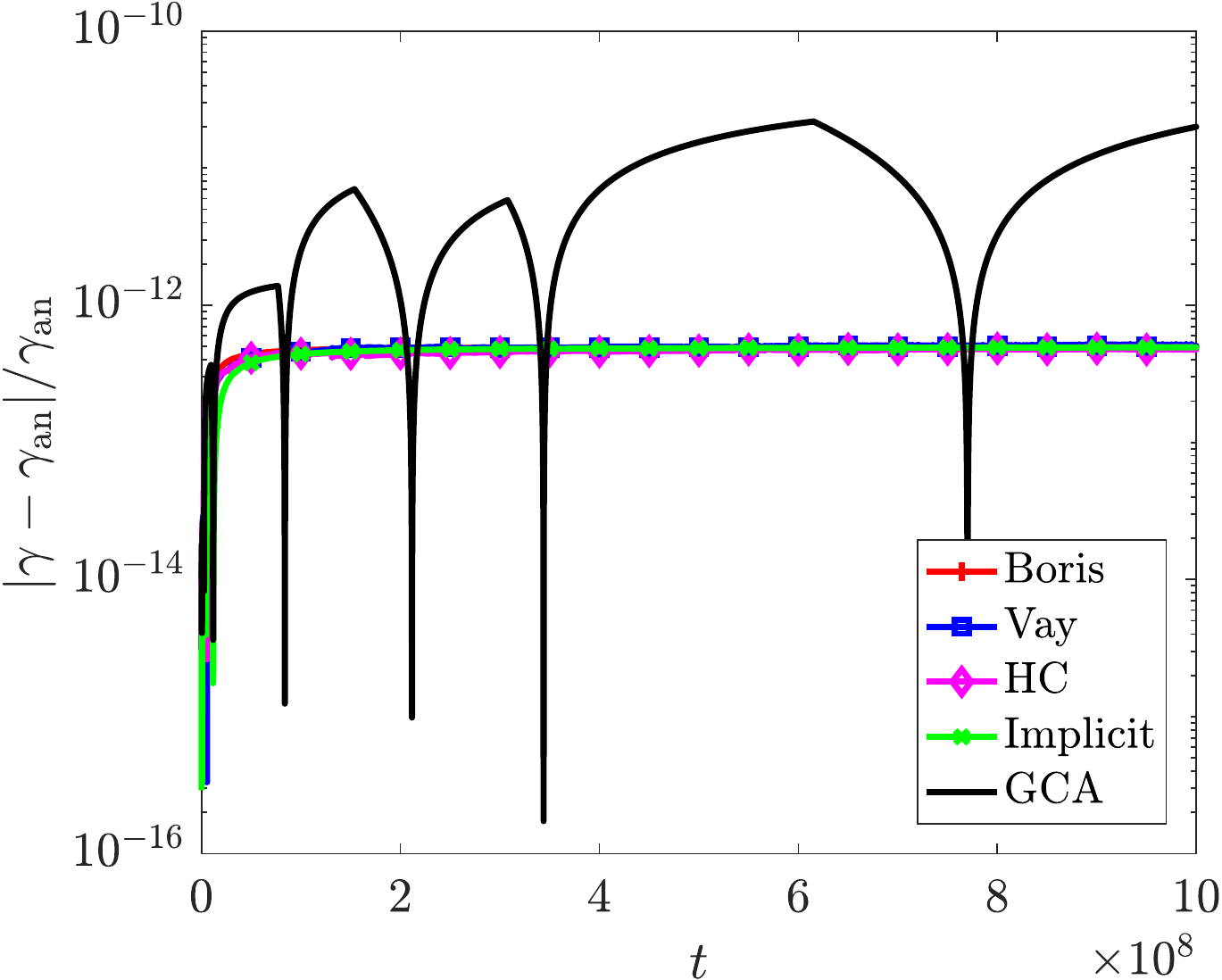}
\caption{Relative error in $\gamma$ for the uniform $\mathbf{E}$ field test. The errors for all methods except GCA are almost indistinguishable.}
\label{fig:acc_errg}
\end{figure}

Fig.~\ref{fig:acc_errg} shows the relative errors measured on all quantities. All the methods perform equally well, calculating the correct Lorentz factor. The apparent deviation of the computed $\gamma$ from the exact value can be safely attributed to truncation relative to finite machine precision, since the error affecting the computed  $v_x$ is of the order of machine precision.

\begin{figure}[!h]
\centering
\includegraphics[width=0.5\columnwidth]{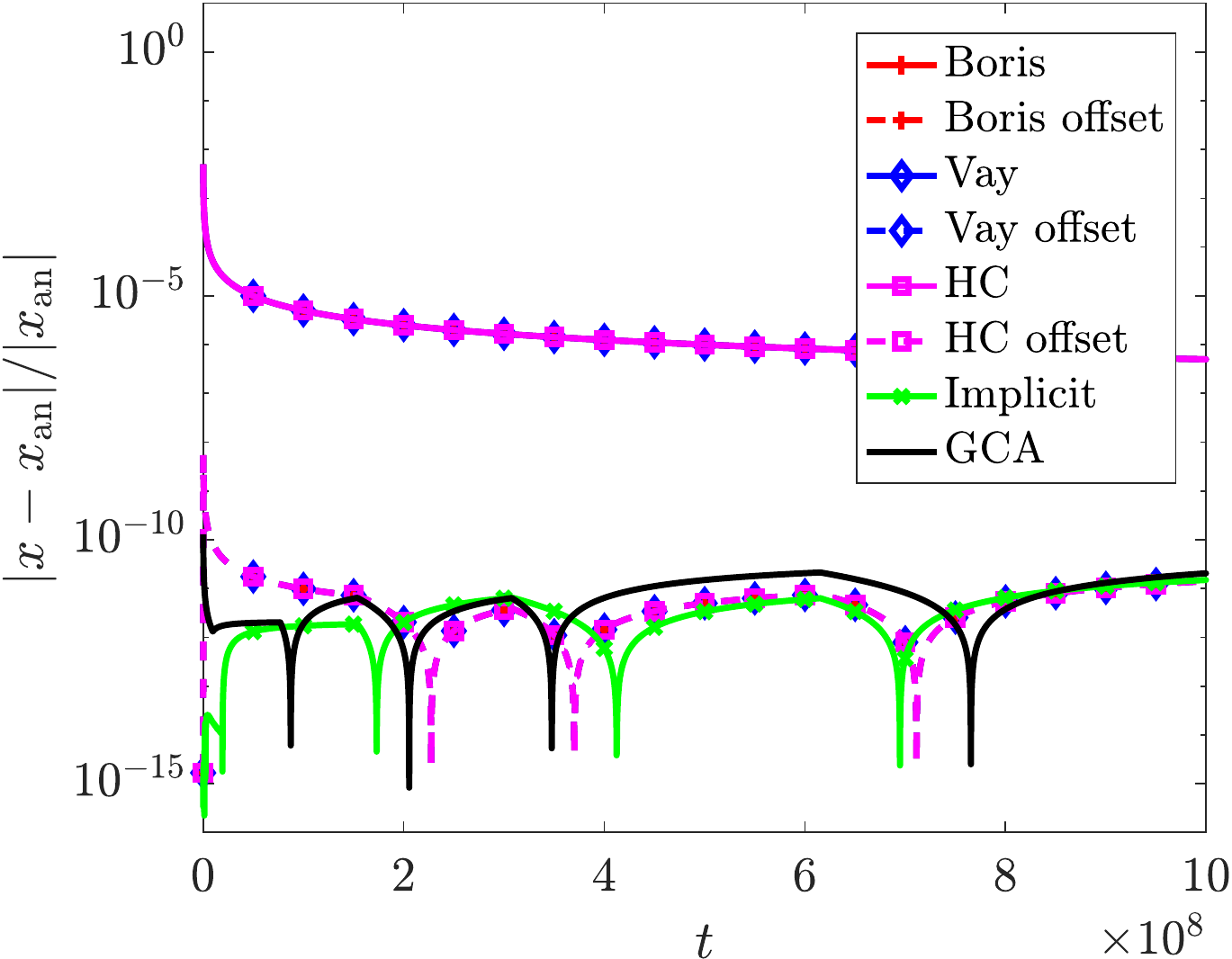}
\caption{Relative error in the position for the uniform $\mathbf{E}$. The results obtained with the Boris, Vay, and HC schemes are shown with (dashed lines) and without (solid lines) initial offset. The results for the three explicit leap-frog schemes almost perfectly overlap for both cases with and without initial offset. The error due to the initial position is pronounced clearly here because within the initial time step a large Lorentz factor is already reached, given the strong electric field.}
\label{fig:acc_errx}
\end{figure}

The error in the position (solid lines in Fig.~\ref{fig:acc_errx}) is above machine precision for the Boris, Vay, and HC schemes, while the implicit method and the GCA perform better. This is an issue that characterizes the relativistic regime, where contrary to the Newtonian equivalent, the evolution of the velocity is nonlinear. Thus, second-order explicit schemes cannot capture the evolution of the position exactly, especially in the initial stages of acceleration (at late times, the velocity is close to the speed of light). Note that, for the same parameters, \cite{Petri2017} observes a much smaller error than with any explicit scheme, while solving the discretized equations with an implicit scheme and the same choice of average velocity as in \cite{Vay_2008}. 

The problem can be mitigated by modifying the synchronized leap-frog scheme as follows. Since the analytic solution is available, we can set the initial ``real'' value $x^{1/2}=x_\mathrm{an}(t=1/2)$, instead of performing a half position update at the very first iteration. This way, the  value of $x^1$ is expected to be closer to the real value. With this modification, we run the test a second time with the explicit schemes and we check for improvements in the computed position.

The results of both runs with and without the modified initial condition are shown in Fig.~\ref{fig:acc_errx}. The error in the position, for runs with modified initial position, is orders of magnitude smaller and comparable to the error from the GCA and implicit results. Thus the initial offset introduced naturally by the leap-frog formulation creates a small displacement in the particle position, leading to a significantly higher error. While using an analytic initial condition solves the problem, it is clear that this is not applicable in practice in a general case when the real solution is not available. Note that the error in the three leap-frog explicit methods is still relatively small. In many applications this could be acceptable compared to the cost of an implicit simulation or a higher order RK scheme such as the one used in the GCA. For mildly relativistic regimes, this error will decrease and in Newtonian regime ($\gamma \rightarrow 1$) it will vanish completely. The error is pronounced clearly here because within the initial time step a large Lorentz factor is already reached. It is also worth noting that, as reported in \cite{Petri2017}, decreasing the time step size might not always have a positive effect, as the larger number of operations will accumulate more second-order errors. 

\subsubsection{Uniform magnetic field}
\label{subsect:gyration}

A particle in a uniform magnetic field, in the absence of electric forces, gyrates on a perfect circle around the guide field line, while conserving its perpendicular velocity $v_\perp$. In the relativistic regime, the gyroradius is given by
\begin{equation}
 R_c=\frac{\gamma m v_\perp}{|q|B},
 \label{eq:gyroradius}
\end{equation}
where $B$ is the magnitude of the guide field. The relativistic gyrofrequency $\omega_c=|q|B/(\gamma m)$ differs from its Newtonian counterpart and decreases as $\gamma$ increases. Since the magnetic field does no work on the particle, $\gamma$ remains constant during the gyration.

We employ the setup presented in \cite{Petri2017}. A single particle is initialized gyrating on the gyroradius $R_c=1$ with $\gamma=10^6$. The initial velocity is $\mathbf{v} = (0,-v_{\perp},0)$, with a guide field $\mathbf{B}=(0,0,B_z)$. If the particle has no velocity parallel to $\mathbf{B}$, the chosen $\gamma$ determines $v_\perp/c=1-\epsilon$, with $\epsilon \simeq 5\times 10^{-13}$. For a particle with charge $q=1$ and mass $m=1$, this requires a magnetic field $B_z/c \sim 10^6$. We follow the circular motion around the guiding center, located at $\mathbf{x}_c=(0,0,0)$, for 100 complete turns. We choose the time step such that each complete gyration of period $T_c=2 \pi\gamma m/(|q|B)$ is resolved with 100 steps. The accuracy of the methods is determined by analyzing how well the computed $\gamma$ (and therefore $R_c$) are conserved. We can also check for errors in the gyration phase $\theta_c$, which is given analytically by
\begin{equation}
 \theta_{c,\mathrm{an}}=-\omega_c t,
\end{equation}
where the minus sign corresponds to our choice of initial conditions. It is expected for the Boris scheme to introduce a small phase lag of order $(\Delta t)^2$ at each time step. The HC scheme should introduce a smaller phase lag of order $(\Delta t)^3$ \citep{HC_2017}.

\begin{figure}[!h]
\subfloat{\includegraphics[width=0.5\columnwidth]{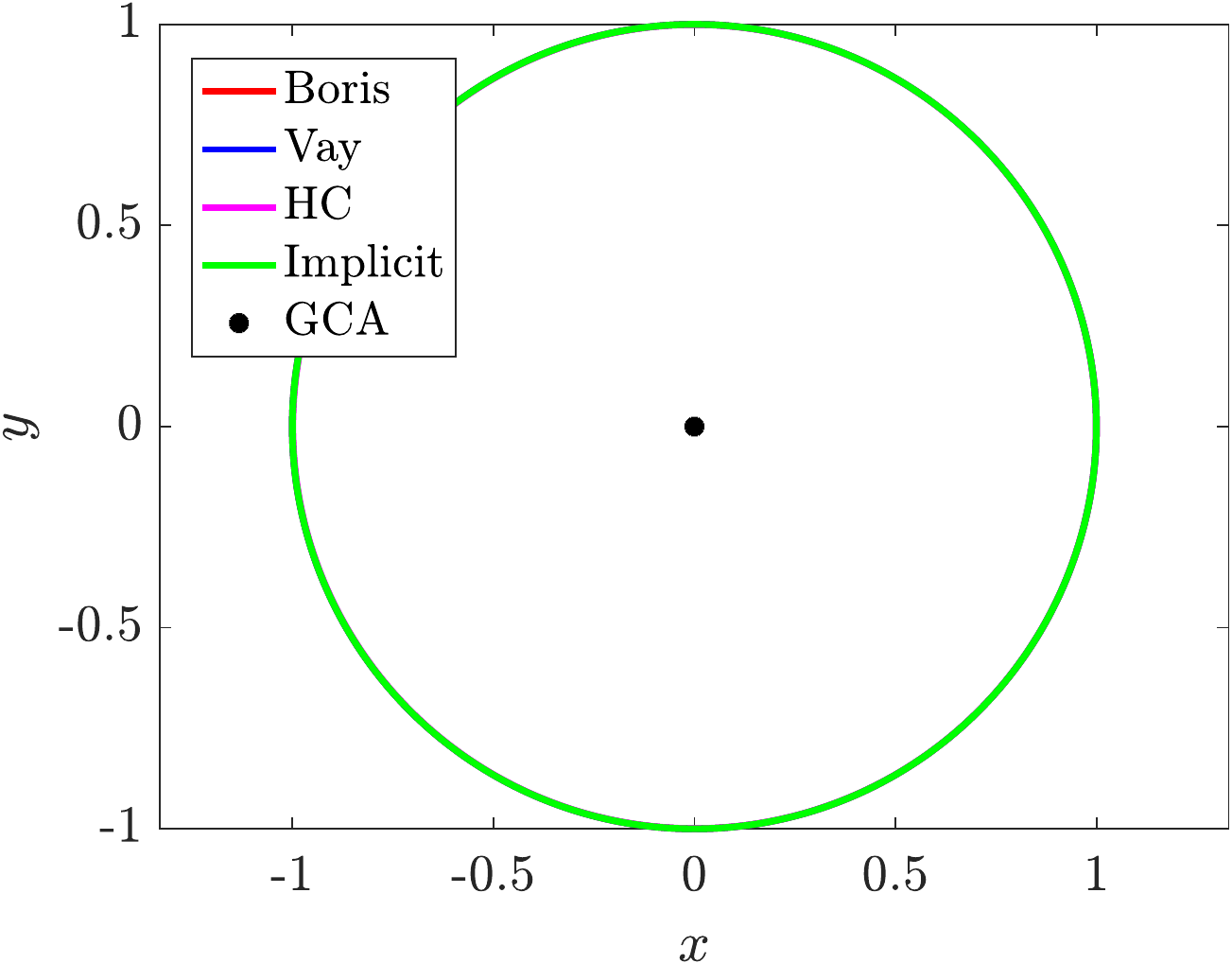}} 
\subfloat{\includegraphics[width=0.5\columnwidth]{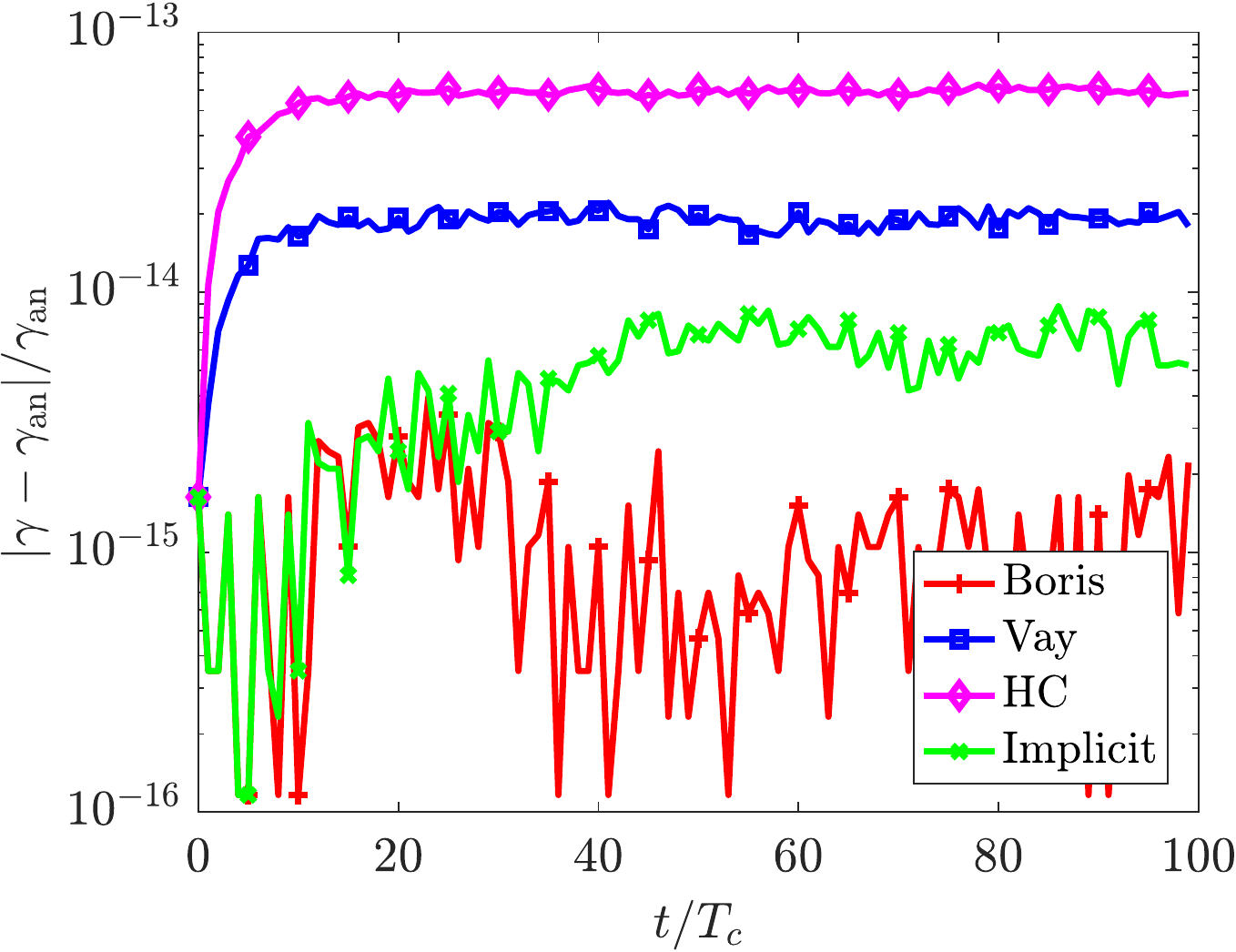}} 
\caption{Results of the uniform $\mathbf{B}$ field test. Left: Trajectory in the $x-y$ plane. The results are visually indistinguishable for all the methods. Right: relative error in $\gamma$.}
\label{fig:gyr}
\end{figure}

Fig.~\ref{fig:gyr} shows the path followed by the particle during the gyration and the conservation of $\gamma$. All the methods correctly confine the particle motion along the circle of radius 1. The GCA result is irrelevant and it is used only as a marker for the position of the guiding center. 

The conservation of $\gamma$ is handled equally well by the Boris and implicit schemes. For Boris, this can be attributed to the way the Lorentz factor is calculated at each time step: if there is no electric field, the same value of $\gamma$ is taken for the magnetic rotation, which in this case corresponds to the exact solution. The HC scheme produces the largest error, while the Vay scheme performs slightly better, but worse than the Boris and the implicit schemes. Contrary to the results of the previous test, these are not pure truncation errors. A deviation from the correct value of $\gamma$, in absence of parallel motion, implies that the perpendicular velocity is varying with respect to the exact (conserved) value. The implicit solution removes the error in $\gamma$. Despite being larger, the error in the Vay and HC schemes is still extremely small and almost of the order of machine precision, a sign that volume preservation is achieved with high accuracy. The error in $\gamma$ directly translates to the error in $R_c$ via Equation \ref{eq:gyroradius}.


\begin{figure}[!h]
\centering
\includegraphics[width=0.5\columnwidth]{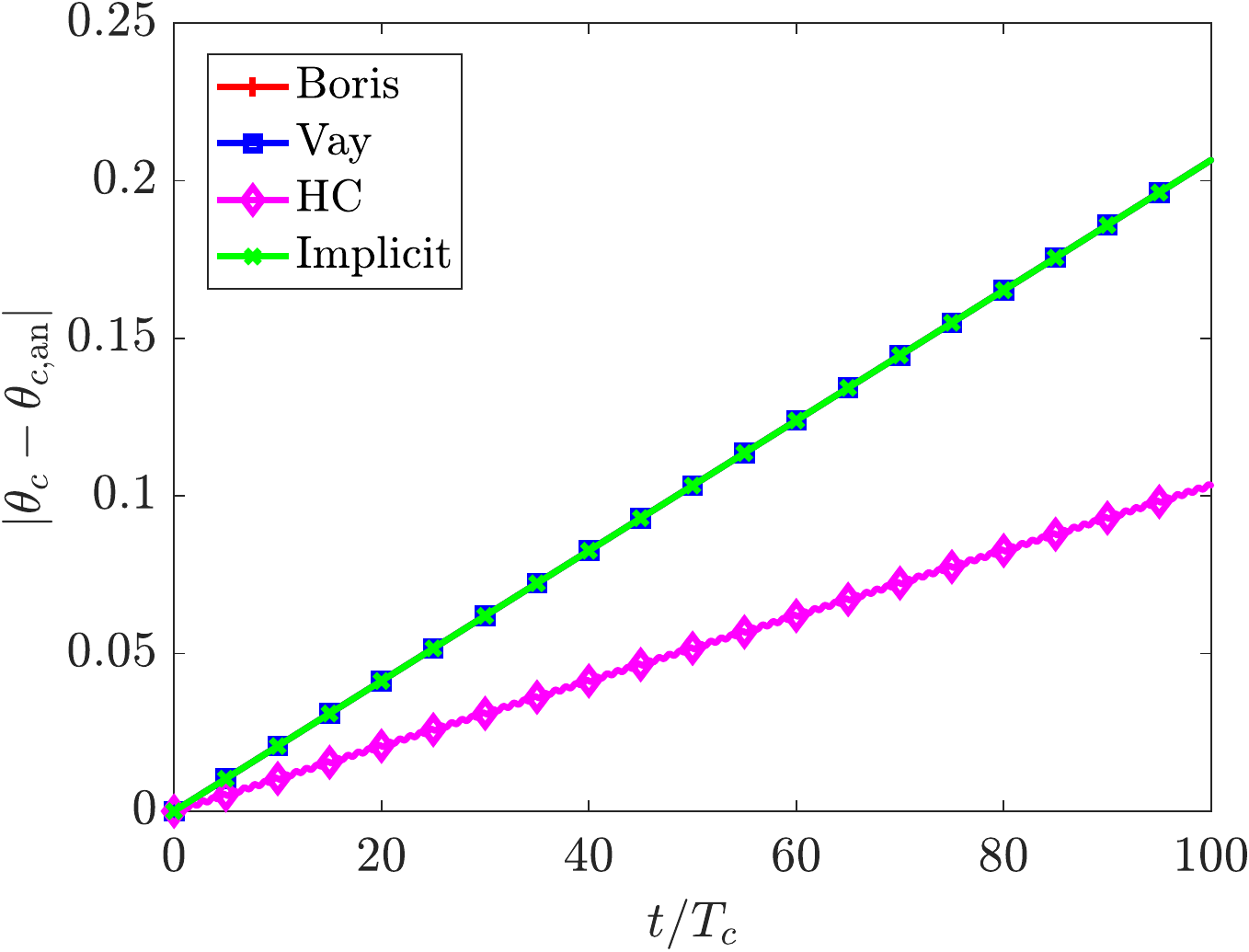}
\caption{Phase lag absolute error for the uniform $\mathbf{B}$ field test. The Boris, Vay, and implicit methods show the same error, while the HC method performs better.}
\label{fig:gyr_errth}
\end{figure}

Fig.~\ref{fig:gyr_errth} shows the phase lag introduced by each method. At each time step, the Boris, Vay, and implicit schemes introduce a small phase lag that accumulates over time. In our case, the gyration is shifted by $\sim 0.2$ radians after 100 turns. The HC scheme produces a smaller phase lag, equal to roughly half of that observed in the other methods, which is compatible with the description of the phase error in the relativistic case as described in \cite{HC_2017}.

\subsubsection{Force-free field}
\label{subsect:FF}

In this section, we present a new test addressing the capabilities of each method in a force-free setup. In the special case $\mathbf{E}=-\mathbf{v}\times\mathbf{B}$, the electric and magnetic forces cancel exactly. The resulting Lorentz force is then
\begin{equation}
\mathbf{F} = q\left(\mathbf{E} + \mathbf{v} \times \mathbf{B}\right) = \boldsymbol{0},
\label{eq:forcefreeLorentz}
\end{equation}
thus there is no evolution in the particle velocity. The particle keeps on traveling at its initial speed with no net change in energy. From the numerical point of view, this test is very stringent, since a slight deviation from exact cancellation of the field forces causes errors in the solution. In the relativistic regime, such errors propagate even more due to the coupling between velocity components through the factor $\gamma$. Note that the force-free condition cannot be obtained for an ensemble of particles with a thermal distribution, but it is still worth analyzing the situation for a single particle.

To test the strength of the schemes, we set up a particle traveling with an initial velocity $v_y$ at $\mathbf{x}=(0,0,0)$. We consider the relativistic regime $\gamma=10^6$, setting up the electric and magnetic fields such that $\mathbf{E}=(E_x,0,0)$ and $\mathbf{B}=(0,0,B_z)$, with $E_x = - v_y B_z$ and $B_z = 1$. The magnitude of the electric field is given by the initial particle velocity, and the force-free condition is ensured. We let the simulation run up to $t=10^5$ with $\Delta t=0.01$ and check for errors in $\gamma$ and the $x$-position, velocity, and momentum, none of which should vary in time.

\begin{figure}[!h]
\subfloat{\includegraphics[width=0.5\columnwidth]{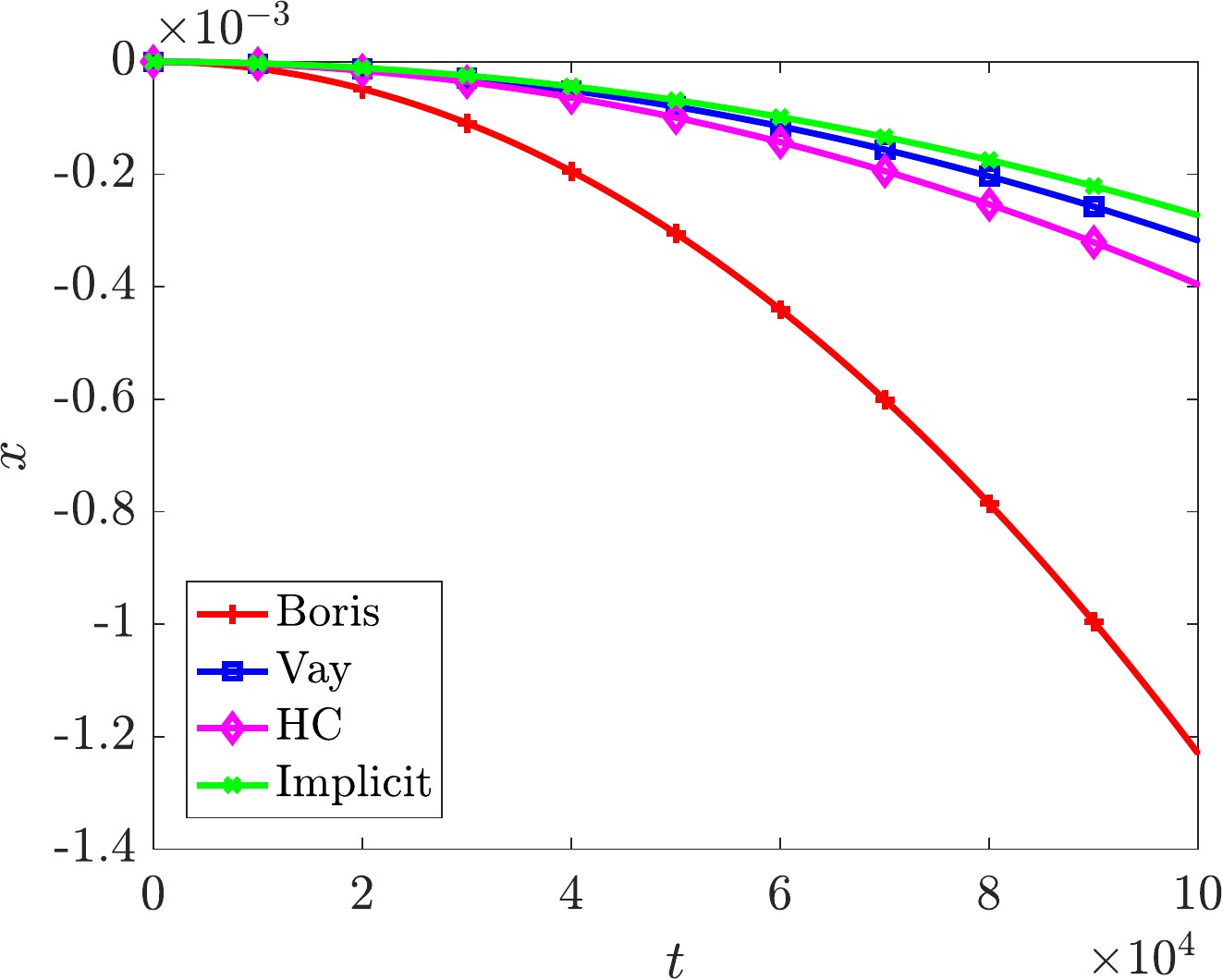}} 
\subfloat{\includegraphics[width=0.5\columnwidth]{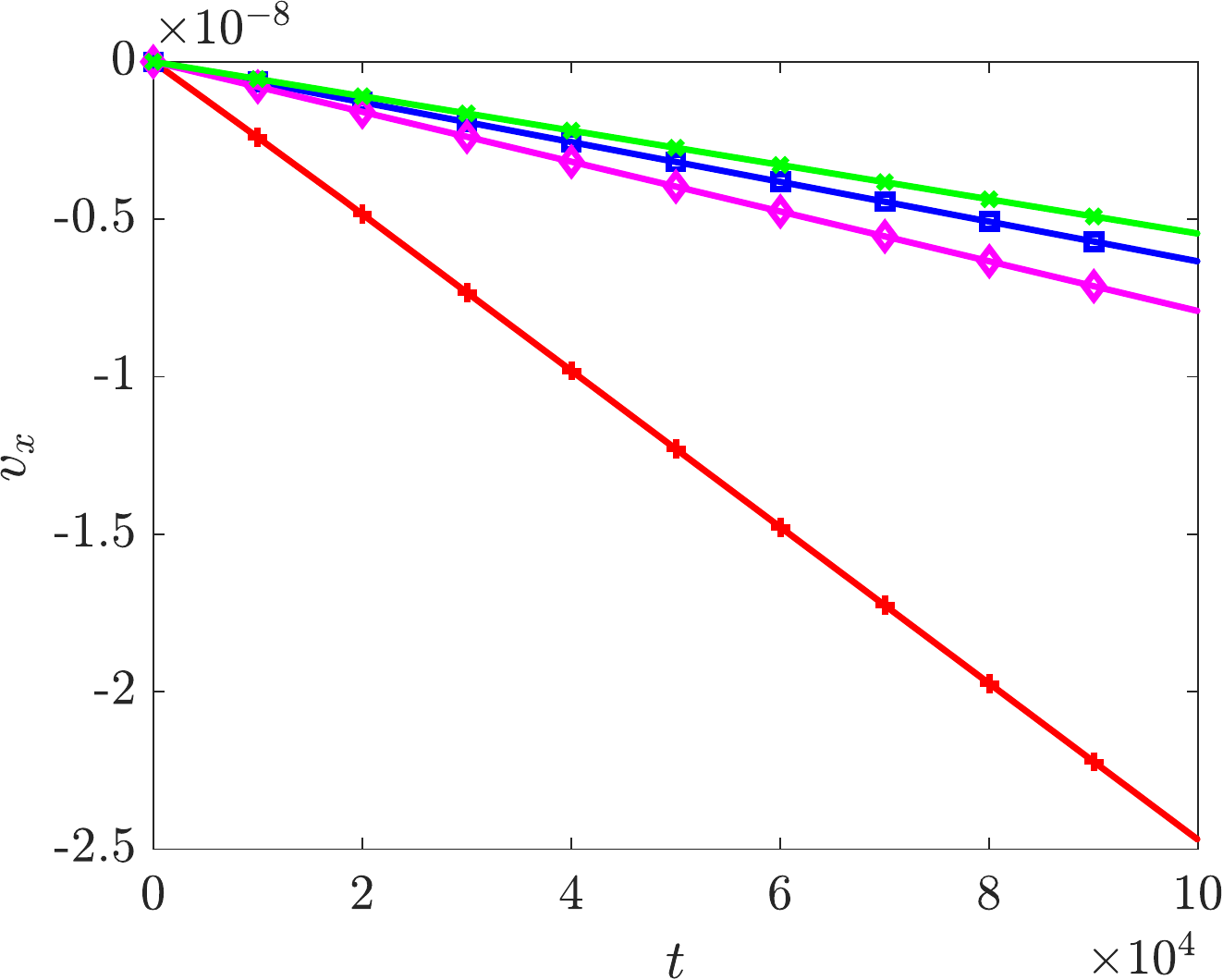}}
\caption{Results of the force-free field test. Left: deviation of the $x-$position from $x_\mathrm{an}=0$. Right: deviation of the $x-$velocity from $v_{x,\mathrm{an}}=0$.}
\label{fig:ff}
\end{figure}

As shown in Fig.~\ref{fig:ff}, all methods eventually deviate from the correct position, velocity, and momentum, with the Boris scheme performing the worst, as predicted by \cite{Vay_2008}. The HC scheme retains better accuracy, close to that obtained with the Vay scheme, which was designed to overcome the Boris scheme limitations in force-free conditions.  The implicit scheme performs better than the others, but still produces spurious deviations from the correct trajectory. For the GCA scheme this is a trivial test, since only the velocity parallel to $\mathbf{B}$ is evolved as a dynamic quantity.

For completeness, we repeat the test by varying the value of $\Delta t$. Thus we can check how the error grows when increasing the time step for the various methods. The results are reported in Table \ref{tab:ff}, where we show the absolute error on the final particle position for each  scheme. The outcome clearly shows that the error for the Boris scheme increases dramatically when increasing $\Delta t$, while for the other methods the growth is much smaller. This is consistent with the properties of the Boris scheme as explained by \cite{Vay_2008}.

\begin{table}
\centering
\begin{tabular}{|c|c|c|c|c|}
\hline
$\Delta t$ & Boris & Vay & HC & Implicit \\ 
\hline 
0.001 & 2.5192$\times 10^{-2}$ & 2.5672$\times 10^{-2}$ & 2.5407$\times 10^{-2}$ & 2.7201$\times 10^{-2}$ \\ 
\hline 
0.01 & 1.2293$\times 10^{-1}$ & 3.1753$\times 10^{-2}$ & 3.9581$\times 10^{-2}$ & 2.7270$\times 10^{-2}$ \\ 
\hline 
0.1 & 4.7705 & 3.9181$\times 10^{-2}$ & 5.1439$\times 10^{-2}$ & 2.7234$\times 10^{-2}$ \\ 
\hline 
1 & 18.7705 & 3.9901$\times 10^{-2}$ & 5.3203$\times 10^{-2}$ & 2.7229$\times 10^{-2}$ \\ 
\hline
\end{tabular} 
\caption{Absolute error on the final particle position along the $x$-axis for the force-free test, for different $\Delta t$. The error affecting the Boris scheme increases approximately one order of magnitude per increasing time step, whereas for the other methods the error does not depend much on the timestep.}
\label{tab:ff}
\end{table}

Interestingly, in our results we observe no error in $\gamma$, meaning that the error in $v_x$ is transferred to $v_y$ with no overall change in the particle energy. This is observed for all runs at different $\Delta t$.

\subsubsection{Perpendicular electric and magnetic fields}
In the specific case where $\mathbf{E} = - \mathbf{v} \times \mathbf{B}$ all forces are canceled, however in typical plasmas perpendicular electric and magnetic fields ($\mathbf{E} \cdot \mathbf{B} = 0$) result in a drifting motion of the particle, perpendicular to both fields. The average motion is in the $\mathbf{E}\times\mathbf{B}$ direction with drift velocity $\mathbf{v}_E = \mathbf{E} \times \mathbf{B}/B^2$. This expression is only valid in the case of weak electric fields ${E}_{\perp} < c B$, with $E_{\perp}$ the electric field perpendicular to the magnetic field. The relativistic drift speed is measured with a Lorentz factor for the drift $\kappa = 1/\sqrt{1-v_E^2/c^2}$. Similar to a test presented by \cite{Petri2017}, we apply an electric field, $\mathbf{E} = (E_0,0,0)$ and a magnetic field $\mathbf{B} = (0,0,1)$, with $E_0$ determining $\kappa$. We choose $E_0/c = 1-\epsilon$, with $\epsilon = 5 \times 10^{-5}$ such that $\kappa = 100$. A particle with $q=1$ and $m=1$ is initalized at the origin $\mathbf{x} = (0,0,0)$ with a velocity $\mathbf{v}=(0,0,0)$. We let the simulation run up to $t = 2 \pi \times 10^7$ with $\Delta t = 0.5$, such that the particle undergoes ten gyrations during its drift. The numerical experiment has also been verified for a particle in an electric field with $E_0/c = 1 - 5 \times 10^{-3}$ such that $\kappa=10$, running up to $t = 2 \pi \times 10^4$ with $\Delta t = 0.0005$.  

\begin{figure}
\subfloat{\includegraphics[width=0.5\columnwidth]{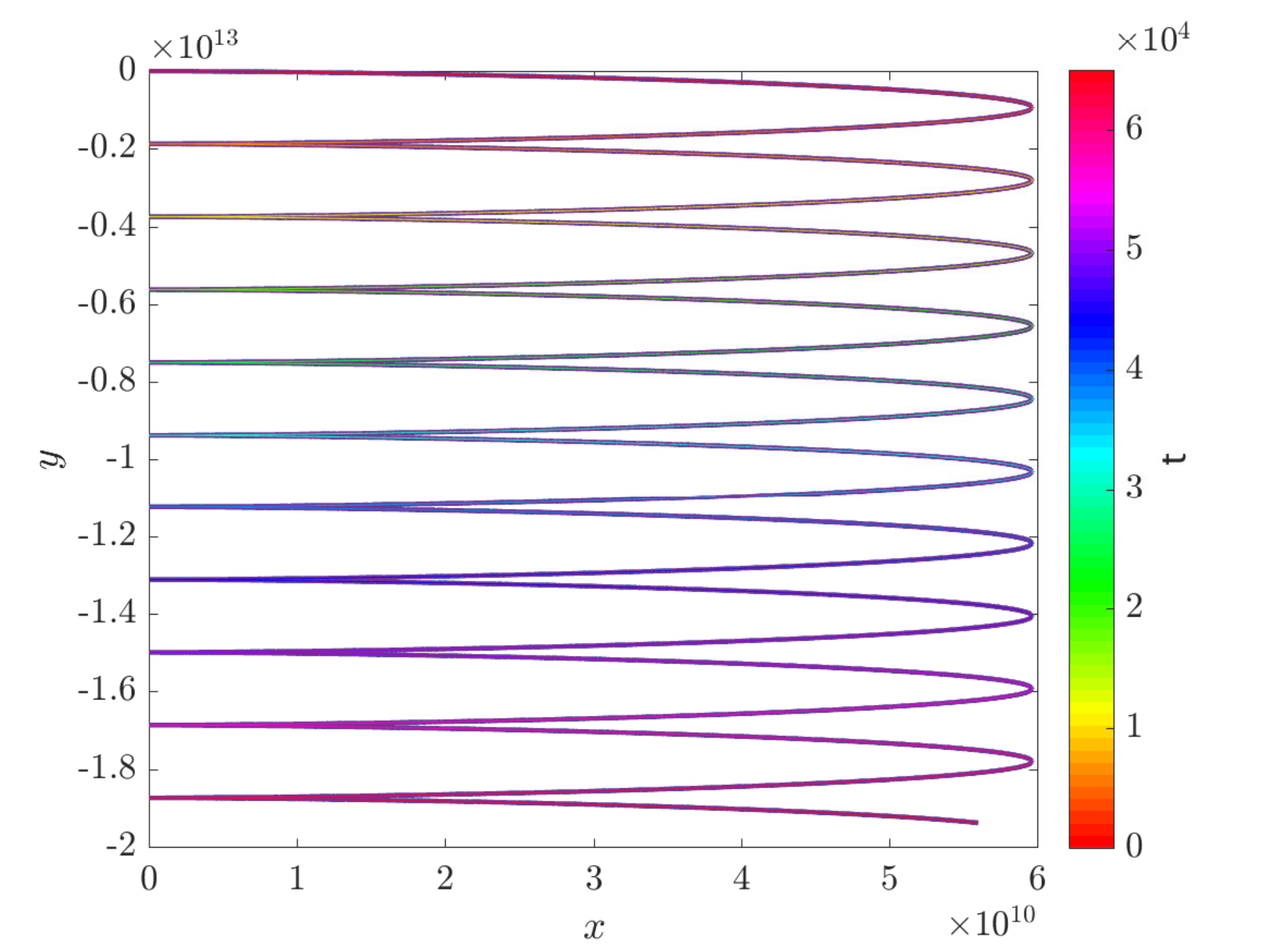}}
\subfloat{\includegraphics[width=0.5\columnwidth]{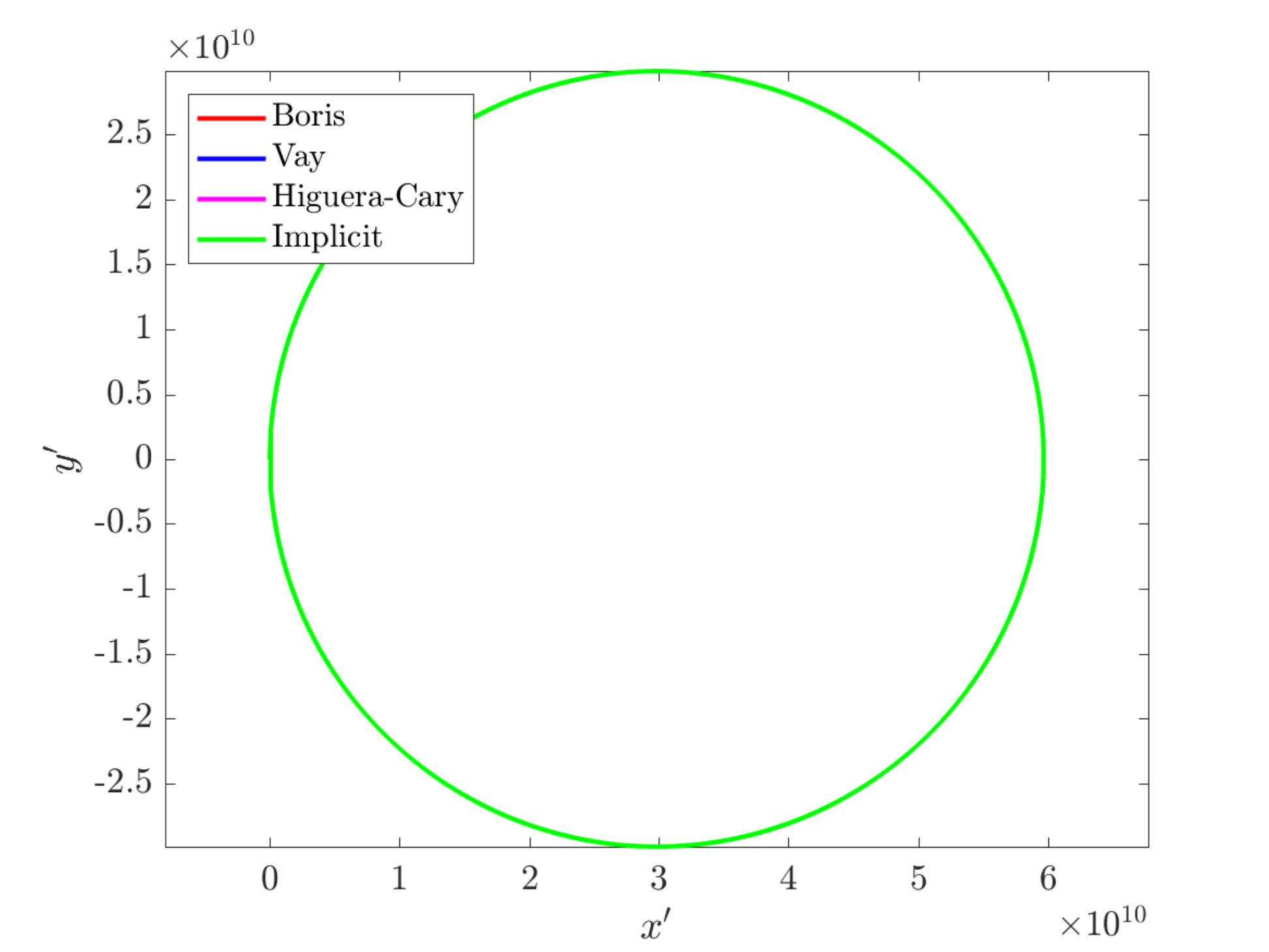}}
\caption{The trajectory of the particle colored by time in the observer frame (left-hand panel) and colored by method, in the comoving $\mathbf{E}\times\mathbf{B}$-frame (right-hand panel) for $\kappa=10$.}
\label{fig:ExB_trajectory_EoverB1}
\end{figure}

\begin{figure}
\subfloat{\includegraphics[width=0.5\columnwidth]{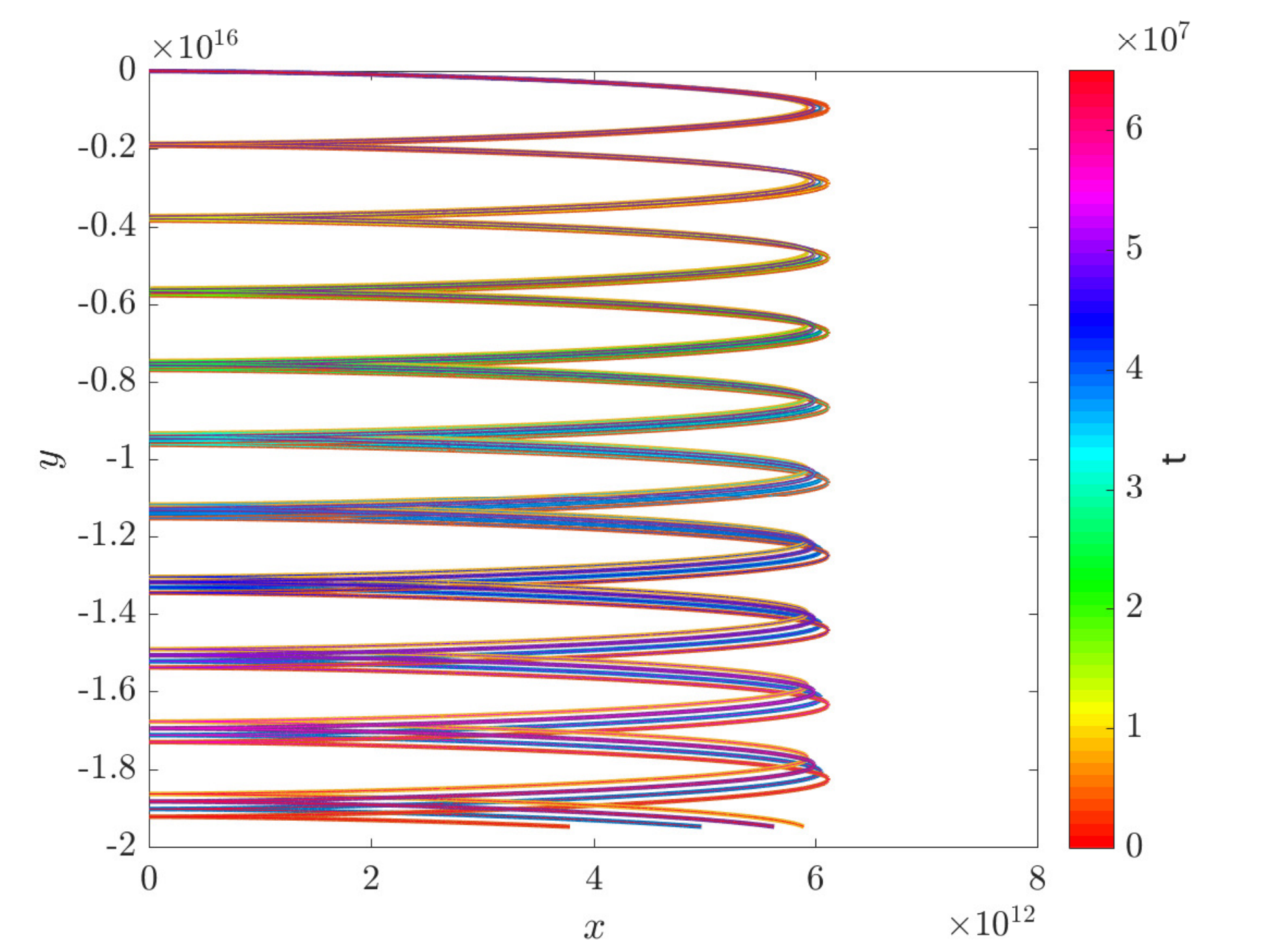}}
\subfloat{\includegraphics[width=0.5\columnwidth]{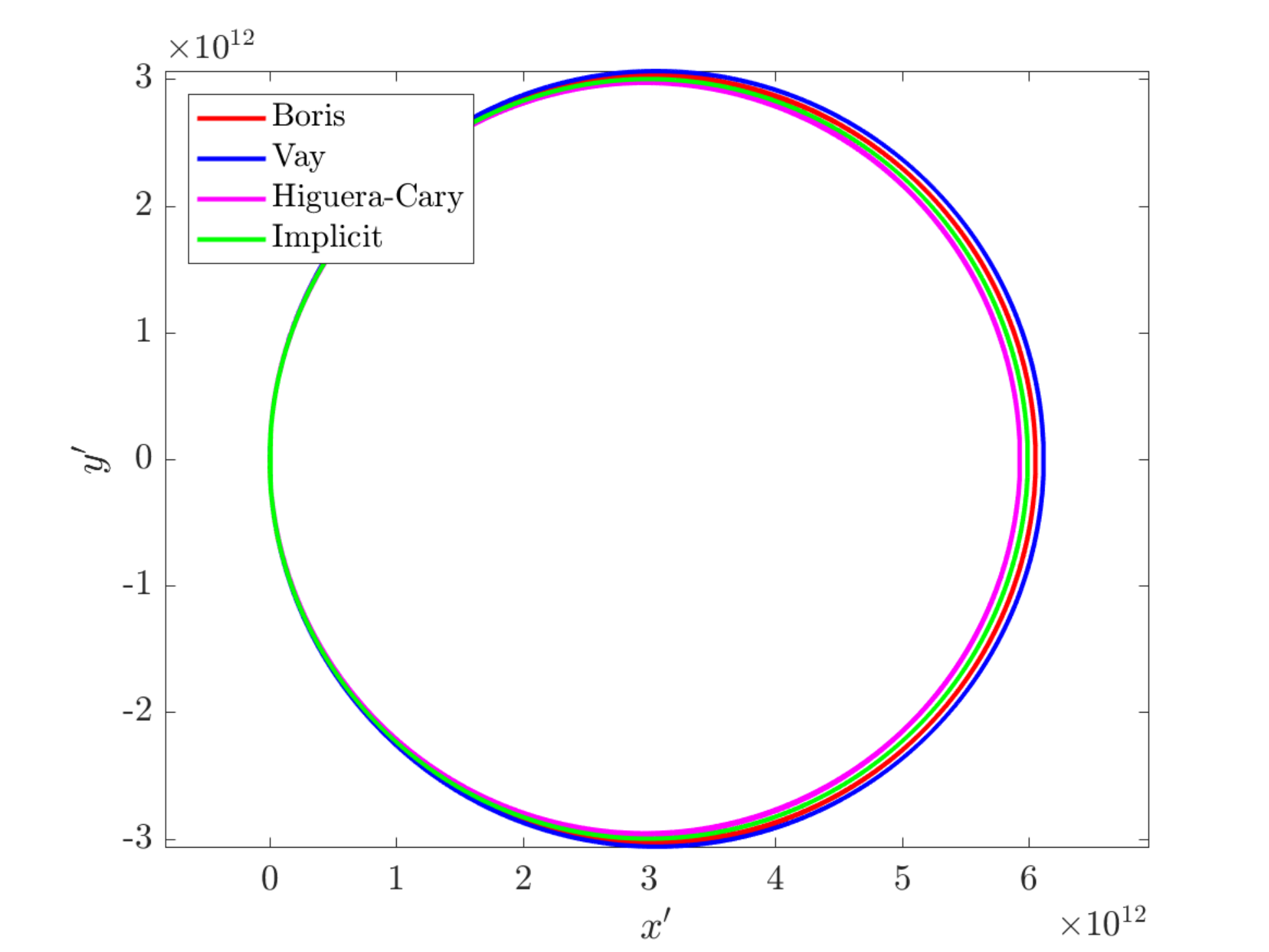}}
\caption{The trajectory of the particle  colored by time in the observer frame (left-hand panel) and colored by method, in the comoving $\mathbf{E}\times\mathbf{B}$-frame (right-hand panel) for $\kappa=100$.}
\label{fig:ExB_trajectory_EoverB2}
\end{figure}

The simulation is conducted in the observer frame, where the particle both drifts and gyrates. We analyze the results both in the observer frame and in the frame comoving with the $\mathbf{E} \times \mathbf{B}$-velocity. Performing a Lorentz boost on the resulting motion, from the observer frame to the $\mathbf{E} \times \mathbf{B}$-frame results in a vanishing electric field and a particle gyrating along the magnetic field. In the comoving frame this results in
\begin{equation}
\mathbf{E}' = \mathbf{0}
\end{equation}
\begin{equation}
\mathbf{B}' = \mathbf{B}/\kappa.
\end{equation}
The coordinates and velocities are boosted to the comoving frame as
\begin{equation}
x' = x,
\end{equation}
\begin{equation}
y'=\kappa(y-v_Et),
\end{equation}
\begin{equation}
z'=z,
\end{equation}
\begin{equation}
v_x' = \frac{v_x}{\kappa}\frac{1}{1-v_E v_y/c^2},
\end{equation}
\begin{equation}
v_y' = \left(\frac{v_y}{\kappa}-v_E+\frac{1}{c^2}\frac{\kappa v_y v_E^2}{\kappa+1}\right)\frac{1}{1-v_E v_y/c^2},
\end{equation}
\begin{equation}
v_z'=v_z,
\end{equation}
resulting in a boosted Lorentz factor
\begin{equation}
\gamma' = \kappa \gamma \left(1 - \frac{v_E v_y}{c^2}\right).
\end{equation} 
The accuracy is measured by the error in the gyroradius in the comoving frame of reference $R_c$ and the error in the comoving Lorentz factor $\gamma'$. Both quantities should be conserved in the comoving frame of reference. The gyroradius is calculated as $R_c = m \gamma' v_{\perp}/q |B'|$. The trajectory of the particle in the observer frame is shown in the left-hand panels of Figures~\ref{fig:ExB_trajectory_EoverB1} and \ref{fig:ExB_trajectory_EoverB2} for $\kappa = 10$ and $\kappa=100$ respectively. The trajectory is colored by time. To distinguish between the four methods we show the trajectory in the comoving frame in the right-hand panels. A slight deviation between the methods is visible for $\kappa=100$, where it has to be noted that a much larger timestep is used here than for the runs with $\kappa=10$. 

\begin{figure}
\subfloat{\includegraphics[width=0.5\columnwidth]{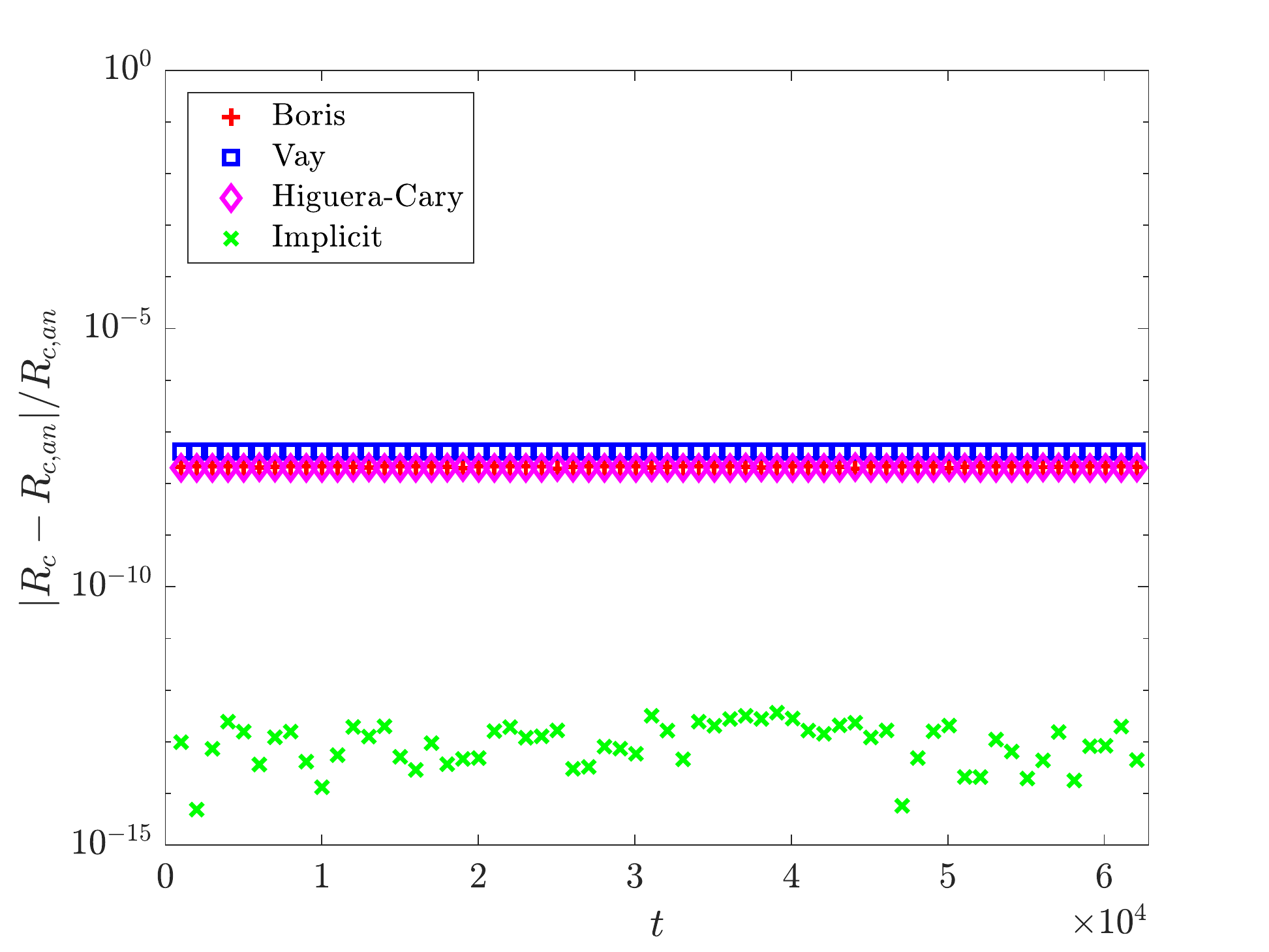}}
\subfloat{\includegraphics[width=0.5\columnwidth]{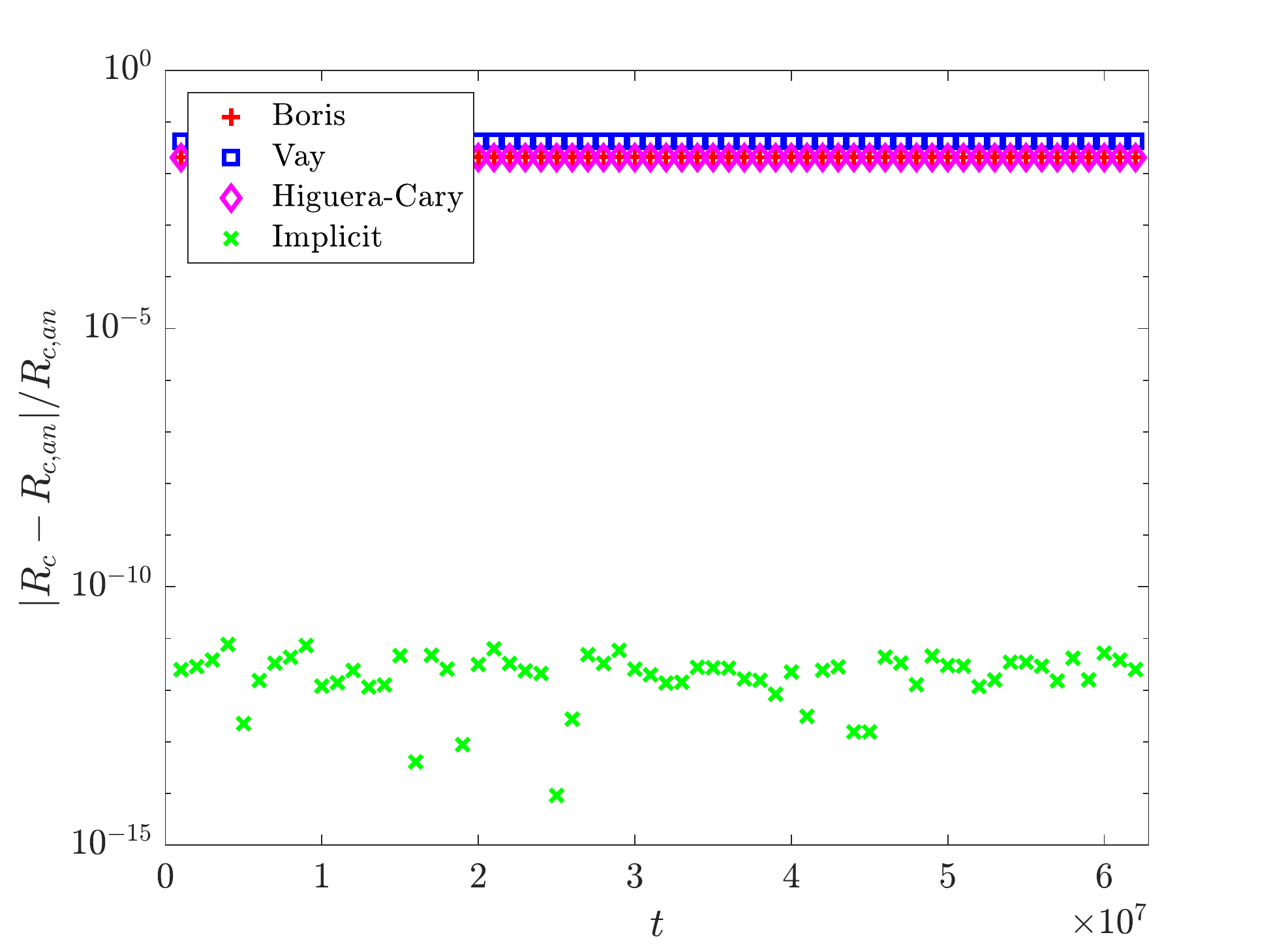}}
\caption{The relative error in the comoving gyroradius (see the right-hand panels of Figures~\ref{fig:ExB_trajectory_EoverB1} and \ref{fig:ExB_trajectory_EoverB2}) for $\kappa=10$ (left-hand panel) and $\kappa=100$ (right-hand panel) calculated by $R_c = m \gamma' v_{\perp}/q |B'|$. The relative errors for the Boris scheme and the HC scheme overlap and are indistinguishable here, whereas they result in a different gyromotion in Fig.~ \ref{fig:ExB_trajectory_EoverB2}.}
\label{fig:ExB_Rcerror_EoverB1}
\end{figure}

From the error in the gyration radius in Fig.~\ref{fig:ExB_Rcerror_EoverB1}, for $\kappa=10$ (left-hand-side) and $\kappa=100$ (right-hand-side), it can be seen that the implicit method gives the correct gyroradius (up to machine precision), whereas all three explicit methods show a nonzero error resulting from the error in the momentum that grows for larger $\kappa$. The error in the gyroradius follows from the error in the Lorentz factor, via $R_c = m \gamma' v_{\perp}/q |B'|$, and not from an error in the position. 

\begin{figure}
\subfloat{\includegraphics[width=0.5\columnwidth]{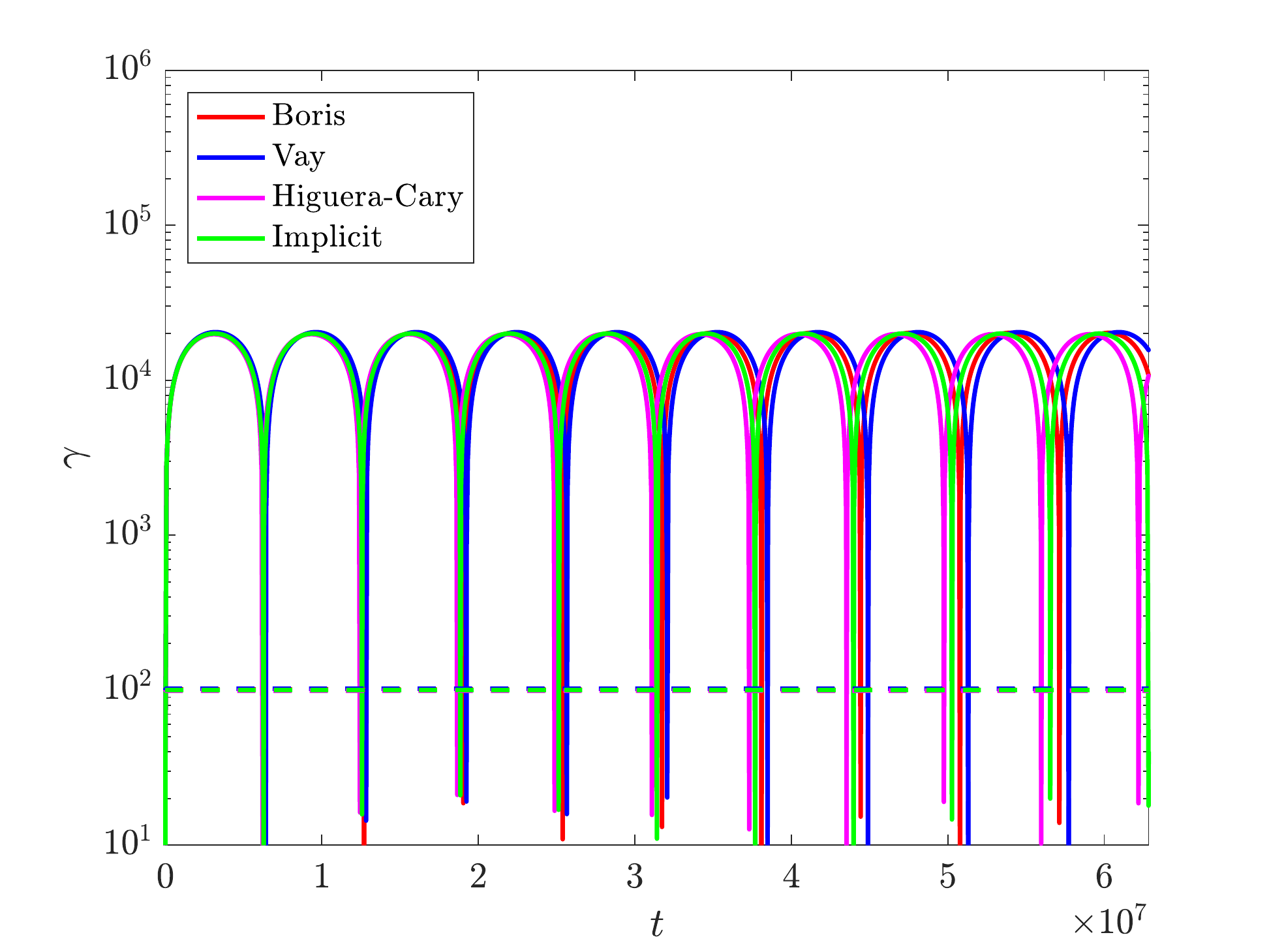}}
\subfloat{\includegraphics[width=0.5\columnwidth]{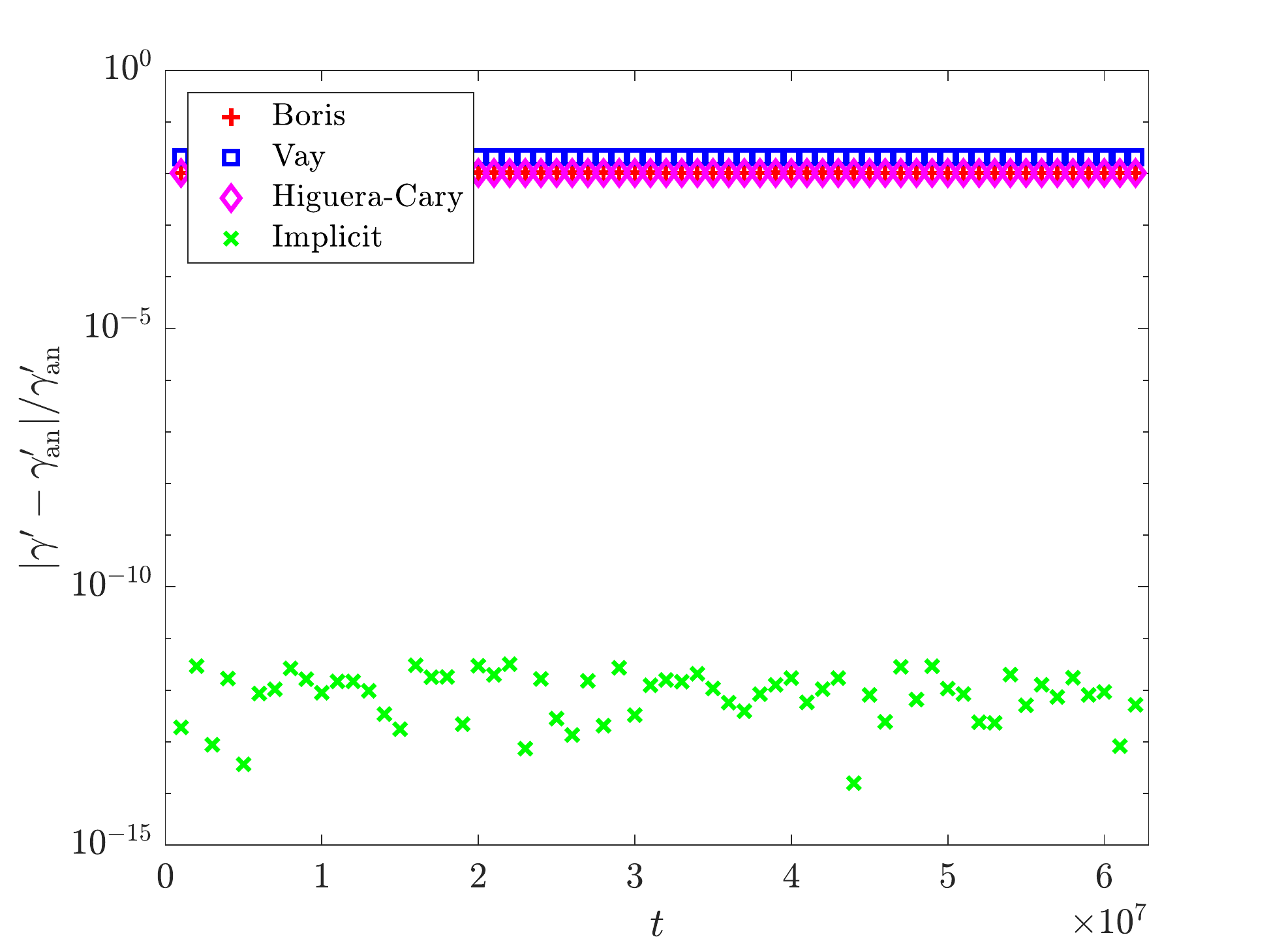}}
\caption{The Lorentz factor in the observer frame (solid lines) and in the comoving frame (dashed lines) in the left-hand panel and the relative error in the comoving Lorentz factor in the right-hand panel for $\kappa=100$. The error for the Boris scheme and the HC scheme overlap in the right-hand panel and are indistinguishable. The error in the gyroradius $R_c = m \gamma' v_{\perp}/q |B'|$ in the right-hand panel of Fig.~\ref{fig:ExB_Rcerror_EoverB1} can be directly related to the difference in the evolution of $\gamma$ in this left-hand panel and to the error in $\gamma$ in this right-hand panel.}
\label{fig:ExB_gamma_EoverB2}
\end{figure}

The Lorentz factor in the observer frame as determined by all four methods is shown with a solid line in the left-hand panel of Fig.~\ref{fig:ExB_gamma_EoverB2} for $\kappa=100$. In the same plot the dashed line represents the boosted Lorentz factor. In the right-hand panel the relative error in the boosted Lorentz factor shows that the implicit method performs best and the Vay scheme performs worst. However, the error in the boosted Lorentz factor is not fluctuating, meaning that it remains constant but suffers from truncation errors due to the large velocities reached and the fact that we are limited by double precision accuracy. For $\kappa=10$ the error in the Lorentz factor is six orders of magnitude smaller than for $\kappa=100$ and the $\mathbf{E}\times\mathbf{B}$-motion for the different methods is visually indistinguishable. For $\kappa=100$ the error in the Lorentz factor results in a different evolution of $\gamma$ for the different methods. 

\subsection{Non-uniform static fields}
\label{sect:testcases_nonuniform}

\subsubsection{Magnetic mirror}
\label{subsect:mirror}

A particle can be trapped inside magnetic mirror (or bottle) configurations, meaning that the magnetic field geometry is such that the field strength increases with position. A particle traveling on a field line entering stronger magnetic fields increases its perpendicular energy as the particle gyrates faster. This increase comes at the expense of the parallel contribution to the kinetic energy since the total energy is conserved (the magnetic field does no work on the particle). The parallel velocity component decreases accordingly and will vanish at a certain point. The particle is then reflected back in the direction it came from, until it reaches the opposite side of the magnetic mirror, where it is reflected again. A typical magnetic field trapping a particle in a magnetic mirror is a quadratic function of the coordinate in the direction of the field plus a radial component,

\begin{equation}
\mathbf{B}(x,y,z) = B_0\left(1+\frac{z^2}{L^2}\right)\hat{\mathbf{z}} + B_r \hat{\mathbf{r}},
\label{eq:mirrorfield}
\end{equation}
with $r = \sqrt{x^2+y^2}$. Assuming cylindrical symmetry ($\partial_{\phi} = 0$ and $B_{\phi} = 0$) for the mirror configuration we can determine the radial component of the magnetic field via the solenoidal constraint ($\nabla \cdot \mathbf{B} = 0$) as (\citealt{Chen})
\begin{align*}
r B_r = -\int\limits^{r}_{0}{r' \frac{\partial B_z}{\partial z} dr'} \approx -\frac{1}{2} r^2 \left(\frac{\partial B_z}{\partial z}\right)_{r=0}\,,
\end{align*}
leading to
\begin{align}
B_r = -r B_0 \frac{z}{L^2} \,.
\end{align}
where we have used that the $z$-component of the magnetic field does not vary much off axis of the magnetic mirror. We obtain the Cartesian components of the field as
\begin{equation}
\mathbf{B}(x,y,z)  = - x B_0 \frac{z}{L^2}  \hat{\mathbf{x}} - y B_0 \frac{z}{L^2} \hat{\mathbf{y}} + B_0\left(1+\frac{z^2}{L^2}\right) \hat{\mathbf{z}},
\label{eq:mirror_cart}
\end{equation}
and to obtain a highly relativistic particle we set $B_0 = 10^6$ and the gradient length $L = 10^7$. The magnetic mirror term corresponds to the last term in the right-hand-side of the GCA momentum equation (\ref{eq:gcastatic2}), $-{\mu_r}\hat{\mathbf{b}}\cdot\nabla\left[B/\kappa \right]/{\gamma}$ or in its Newtonian limit (\ref{eq:gcanewton2}), $-\mu\hat{\mathbf{b}}\cdot\nabla B$. The latter simplifies in the case of a mirror in the $z$-direction and translates to an evolution equation for $v_{\|} = v_z$, averaged over a gyration (\citealt{Chen})
\begin{equation}
\frac{d m v_z}{dt} = -\mu \frac{\partial B_z}{\partial z}.
\label{eq:mirror_force}
\end{equation}
We recognize the field aligned restoring force, pointing towards the center of the magnetic mirror, opposite to the direction of increasing field strength. Particles with a purely parallel velocity (or a negligible pitch angle) have no magnetic moment $\mu$ and hence do not undergo a bouncing motion. These particles escape from the magnetic mirror, resulting in a loss cone of particles. We can obtain a condition for a particle to mirror by substituting magnetic field (\ref{eq:mirror_cart}) in equation (\ref{eq:mirror_force}), resulting in a mirror length of (\citealt{Bittencourt})
\begin{equation}
z_{max} = \pm L\sqrt{\frac{B_{max}}{B_0}-1}= \pm10^7.
\label{eq:zmax}
\end{equation}
However, this position depends on the assumption that the magnetic moment is conserved. The magnetic moment is an adiabatic invariant, that is only conserved to a certain extent depending on the small parameter $\epsilon = R_c/L$. 

\begin{figure}[!h]
\centering
\includegraphics[width=0.5\columnwidth]{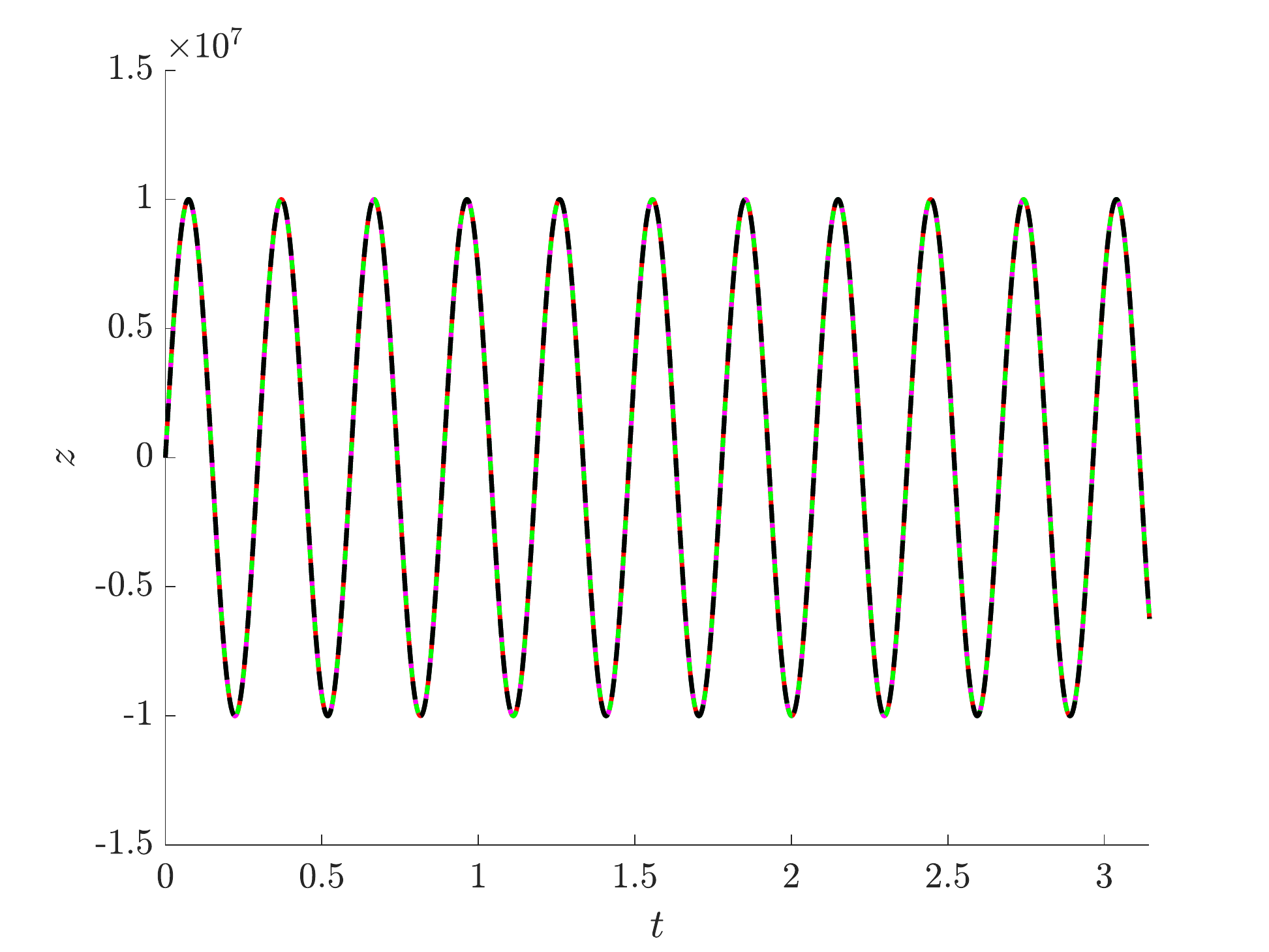}
\caption{The particle trajectory in time on the axis of a magnetic bottle. There is no observable difference after ten cycles through the magnetic bottle between all methods and the GCA results, here shown for the run with interpolated fields and time step $\Delta t = 10^{-8}$ for all methods, except for the Vay method, where we had to use $\Delta t = 10^{-9}$ for the particle not to escape the magnetic bottle. For all methods the particle stays within the analytically predicted range $z_{max} = \pm 10^{7}$.}
\label{fig:mirror_trajectory}
\end{figure}

\begin{figure}
\subfloat{\includegraphics[width=0.5\columnwidth]{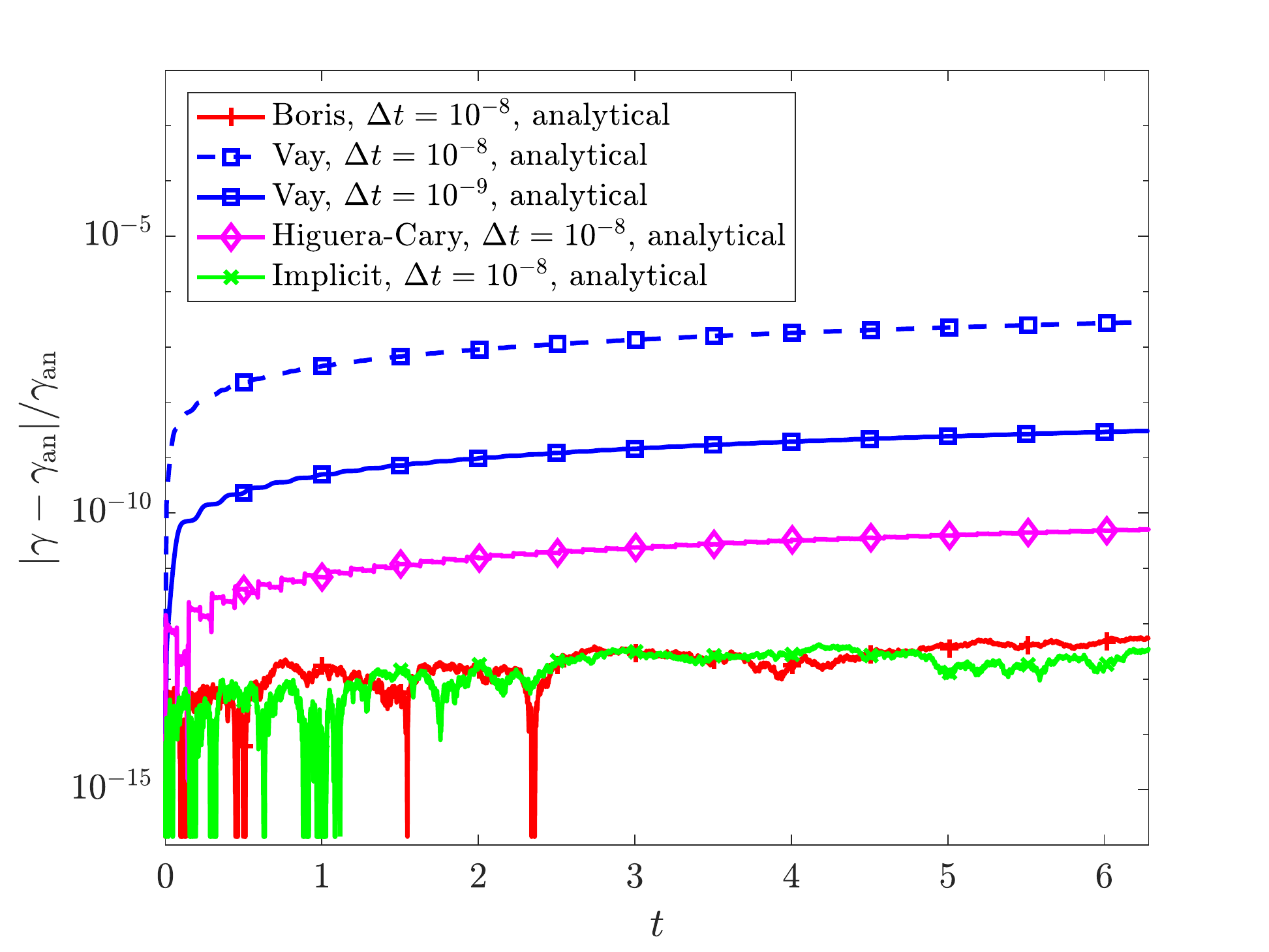}}
\subfloat{\includegraphics[width=0.5\columnwidth]{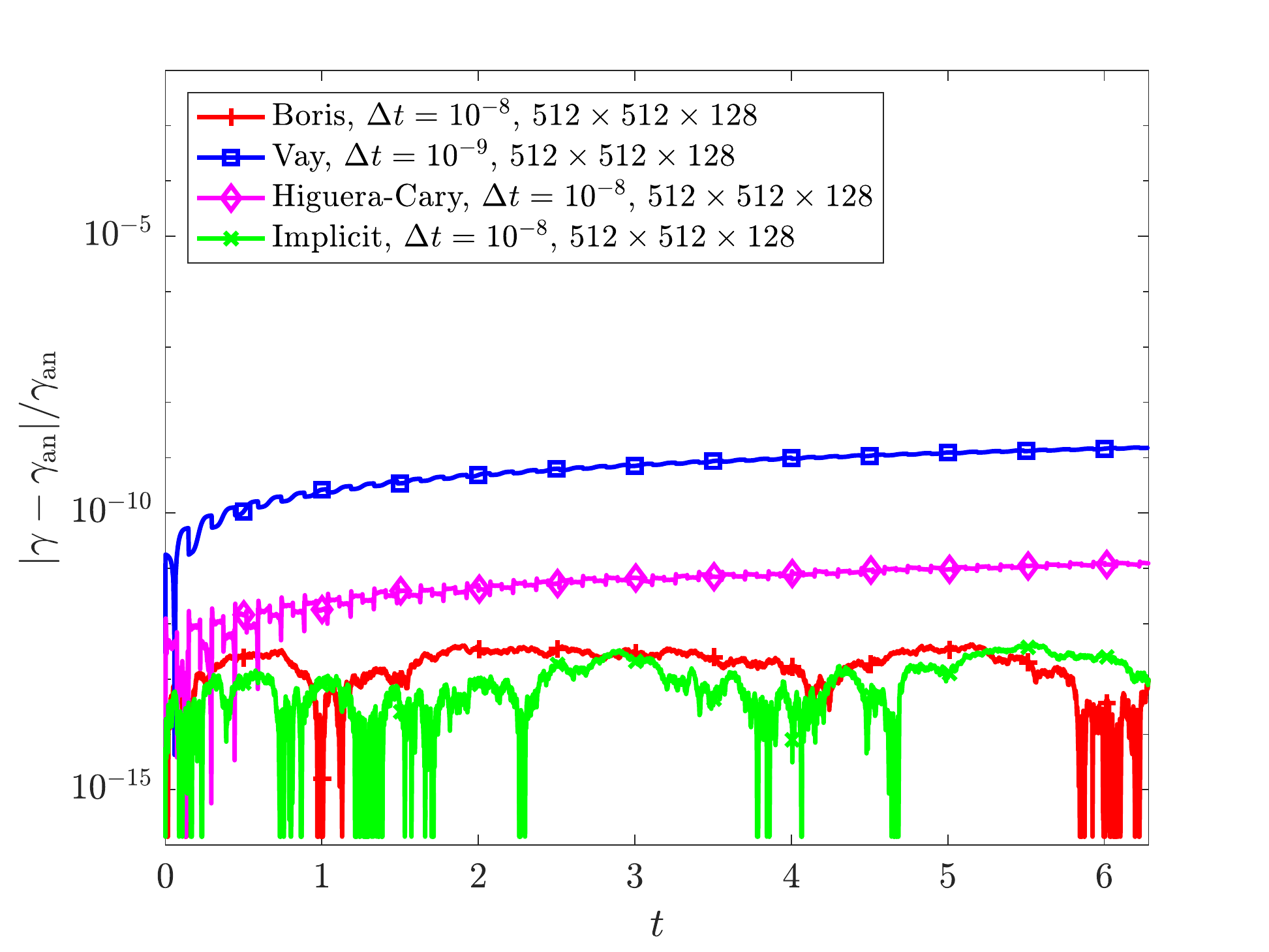}}
\caption{The relative error in the Lorentz factor with analytically given fields (left-hand panel) and interpolated fields (right-hand panel), for ten cycles through the magnetic bottle. The error for the Vay scheme in interpolated fields and time step $\Delta t = 10^{-8}$ is not shown in the right-hand panel because the particle's magnetic moment is not conserved due to numerical errors and the particle escapes the magnetic bottle immediately. }
\label{fig:mirror_gammaerror}
\end{figure}

We initialize a particle with $q=1$ and $m=1$ at $\mathbf{x} = (-R_c, 0, 0)$ with velocity $\mathbf{v} = (0, v_{\perp}, v_{\|})$ with $v_{\perp} = v_{\|} = 0.707c$ such that $\gamma = 100$ and the initial gyroradius is $R_c =\gamma m v_{\perp}/qB_0= 0.0000707c$. The tests have been performed with time steps $\Delta t = 10^{-7}$ , $\Delta t = 10^{-8}$ and $\Delta t = 10^{-9}$. The time step is decreased until the error converges such that it does not differ after taking a smaller time step. We ran with both interpolated fields and analytical fields (except for the guiding center approximation, where we always use interpolation) to rule out any effect of interpolation errors. The test with $\Delta t = 10^{-8}$ has been performed with interpolated fields with a grid resolution of $512\times512\times128$ in a domain $[-20L, 20L] \times [-20L, 20L] \times [-2 \times 10^6L, 2 \times 10^6L]$, and convergence has been confirmed for finer resolutions. There is no visually observable difference between the trajectory in time of the particle along the mirror axis (see Fig.~\ref{fig:mirror_trajectory}) between the explicit methods, the implicit method and the result from the GCA. The trajectories obtained by all methods satisfy the maximum mirror length in Equation (\ref{eq:zmax}).
The accuracy is determined by the relative error in the Lorentz factor in the observer frame, that has to be conserved since there are no electric fields. The Lorentz factor is conserved up to machine precision by both the Boris scheme and the implicit scheme, regardless of whether the fields are given analytically or interpolated (see the left-hand panel of Fig.~\ref{fig:mirror_gammaerror} for the relative error with analytic fields and the right-hand panel for the relative error with interpolated fields). The Higuera-Cary scheme has an error that grows initially but settles to a constant value slightly larger than machine precision. The relative error in $\gamma$ for the Vay scheme shows a similar trend as for the Higuera-Cary scheme, however it grows to a larger value than the error for the HC scheme, even for $\Delta t = 10^{-9}$. For a smaller time step the error does not decrease any more. The error for the Vay scheme in interpolated fields and time step $\Delta t = 10^{-8}$ is not shown because the particle's magnetic moment is not conserved due to numerical errors and the particle escapes the magnetic bottle immediately. 

\begin{figure}
\subfloat{\includegraphics[width=0.5\columnwidth]{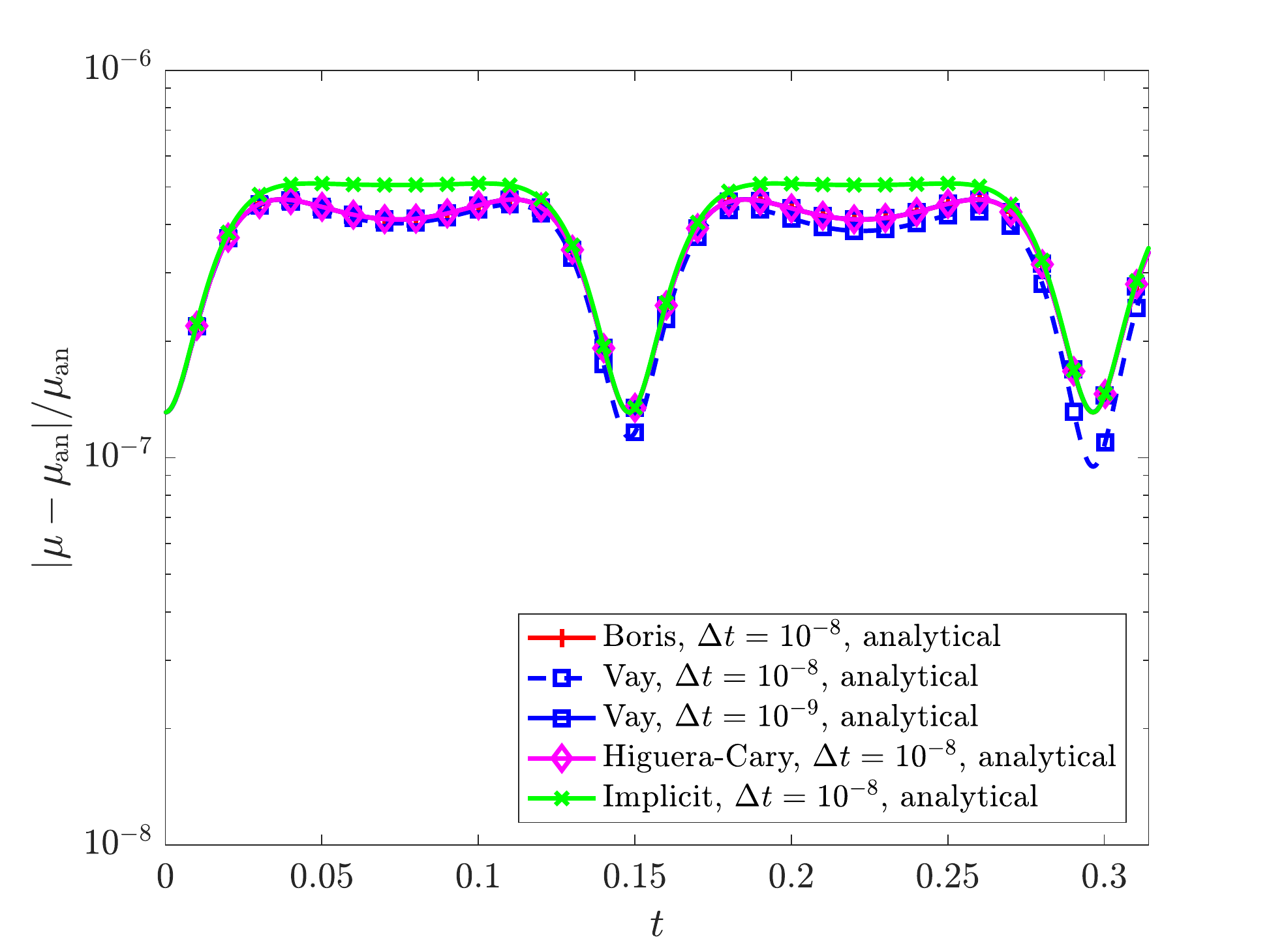}}
\subfloat{\includegraphics[width=0.5\columnwidth]{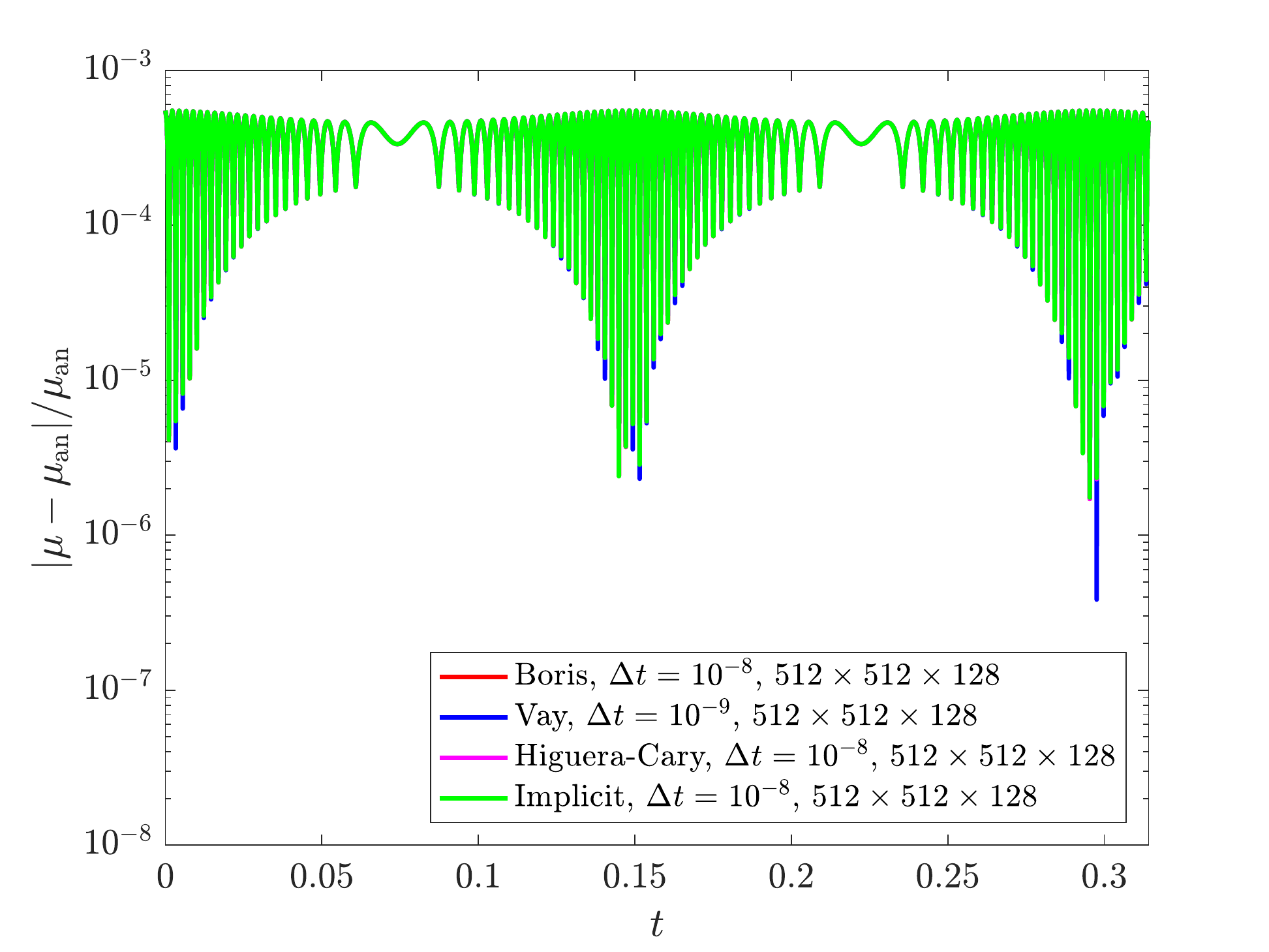}}
\caption{The relative error in the magnetic moment with analytically given fields (left-hand panel) and interpolated fields (right-hand panel) for one cycle in the magnetic bottle. The error for the Vay scheme in interpolated fields and time step $\Delta t = 10^{-8}$ is not shown in the right-hand panel because the particle's magnetic moment is not conserved due to numerical errors and the particle escapes the magnetic bottle immediately. In the left-hand panel, for analytic fields (see the dashed blue line), the error already slightly differs after one cycle, eventually growing unboundedly until the particle escapes after a few cycles. The errors with interpolated fields (right-hand panel) are nearly identical for all methods, and much worse than for analytical fields.}
\label{fig:mirror_muerror}
\end{figure}

We also show the error in the magnetic moment for analytic fields in the left-hand panel of Fig.~\ref{fig:mirror_muerror} and for interpolated fields in the right-hand panel for a fraction of the simulation up to $t = \pi/10$, corresponding to one full cycle through the magnetic bottle. It is harder to draw conclusions from this since $\mu$ is an adiabatic invariant, meaning that conservation is only approximately valid for spatially (and temporally) slowly varying fields. This is the case for $\epsilon = R_c / L \ll1$. In our simulations $\epsilon \approx 0.002$. For larger $\epsilon$ the error in $\mu$ grows. We do observe that the Vay scheme needs a time step that is an order of magnitude smaller than the other methods to reach the same accuracy. If we analyze the relative error in $\mu$ for the full simulation time (10 cycles through the magnetic bottle) we conclude that $\mu$ is conserved less well by the Vay scheme. This results in the particle gaining parallel velocity and losing perpendicular velocity per cycle and eventually the particle will leave the magnetic bottle. For interpolated fields the error in $\mu$ is larger than for analytic fields, whereas for the Lorentz factor this error is of similar order. This shows that the grid resolution affects the efficiency of the mirror and for a coarser grid a particle will end up in the loss cone at an earlier time. For the guiding center approximation, the error in $\mu$ is equal to zero by definition.

\subsection{Tests for the guiding center approximation}
\label{sect:testcases_gca}

With the next three test cases we investigate the accuracy of the GCA. How well the GCA method predicts the trajectory of a gyrating particle depends on the spatial variation of the magnetic field. If
during a gyration $\mathbf{B}$ changes significantly, the approximations employed
in GCA will not be accurate. The relative change in magnetic field can be
expressed as $\delta B/B$, where $\delta B$ is the variation over one gyration,
and $B$ is the magnetic field at the center of gyration. An estimate for
$\delta B/B$ is
\begin{equation}
  \delta B/B \sim \frac{R_c \, |\nabla B|}{B},
\end{equation}
where $\nabla B$ is the gradient of $|\mathbf{B}|$. The tests described below are
for simplicity performed in the non-relativistic regime ($\gamma \approx 1$).
For relativistic particles the validity of GCA still depends on $\delta B/B$,
but then $\delta B/B$ will depend on $\gamma$ since $R_c \propto \gamma$. We
compare the GCA approach with Boris method, leaving out the other particle
movers described in section \ref{sect:numerical}. The reason for this is that
for $\gamma \to 1$, the different movers (Boris, Vay, Higuera-Cary) reduce to
the same scheme, and that for sufficiently small $\Delta t$ the implicit scheme
converges to the same results.

\subsubsection{Magnetic field gradient}
\label{subsect:gradB}

We now consider a perpendicular gradient in the magnetic field strength
\begin{equation}
\boldsymbol{B}(x,y,z) = B_0\left(1+\frac{x}{L}\right)\boldsymbol{\hat{z}},
\label{eq:gradBfield}
\end{equation}
and no electric field ($\boldsymbol{E} = \boldsymbol{0}$). Assuming $R_c \ll L$,
the drift due to such a gradient can be approximated analytically, see e.g.
\citep{Bittencourt}
\begin{equation}
\mathbf{v}_{\nabla B} = \pm \frac{v_{\perp} R_c}{2} \frac{\mathbf{B}\times\nabla B}{B^2}
\label{eq:gradBvel}
\end{equation}
where the $\pm$ depends on the sign of the charge of the particle, being
positive for positive charges. For the field given by equation
\eqref{eq:gradBfield}, and assuming $x > -L$, this reduces to
\begin{equation}
  \mathbf{v}_{\nabla B}
  = \pm \frac{v_{\perp} R_c B_0^2}{2LB^2}\left(1+\frac{x}{L}\right)\boldsymbol{\hat{y}}
  = \pm \frac{v_{\perp} R_c B_0}{2LB}\boldsymbol{\hat{y}}
\label{eq:gradBvel2}
\end{equation}
The direction of the drift is perpendicular to
both the magnetic field and the direction of its gradient, so the momentum
equation yields $d m v_{\|}/dt = 0$ in absence of an electric field.

\begin{figure}
  \centering
  \subfloat{\includegraphics[width=0.5\columnwidth]{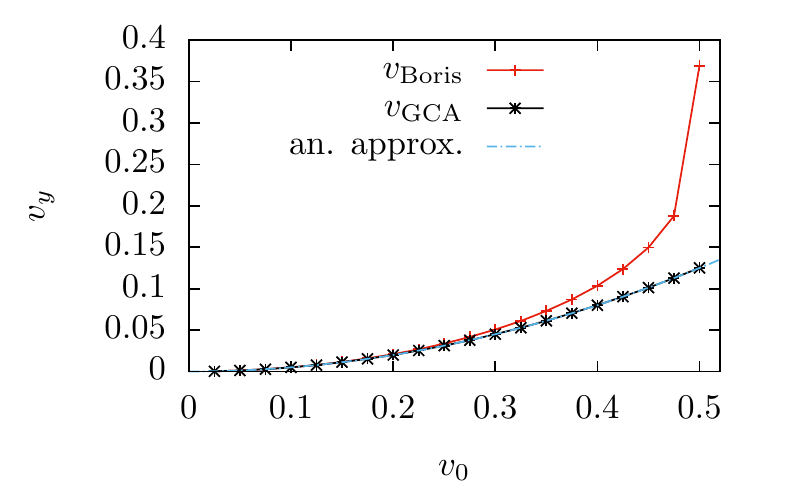}}
  \subfloat{\includegraphics[width=0.5\columnwidth]{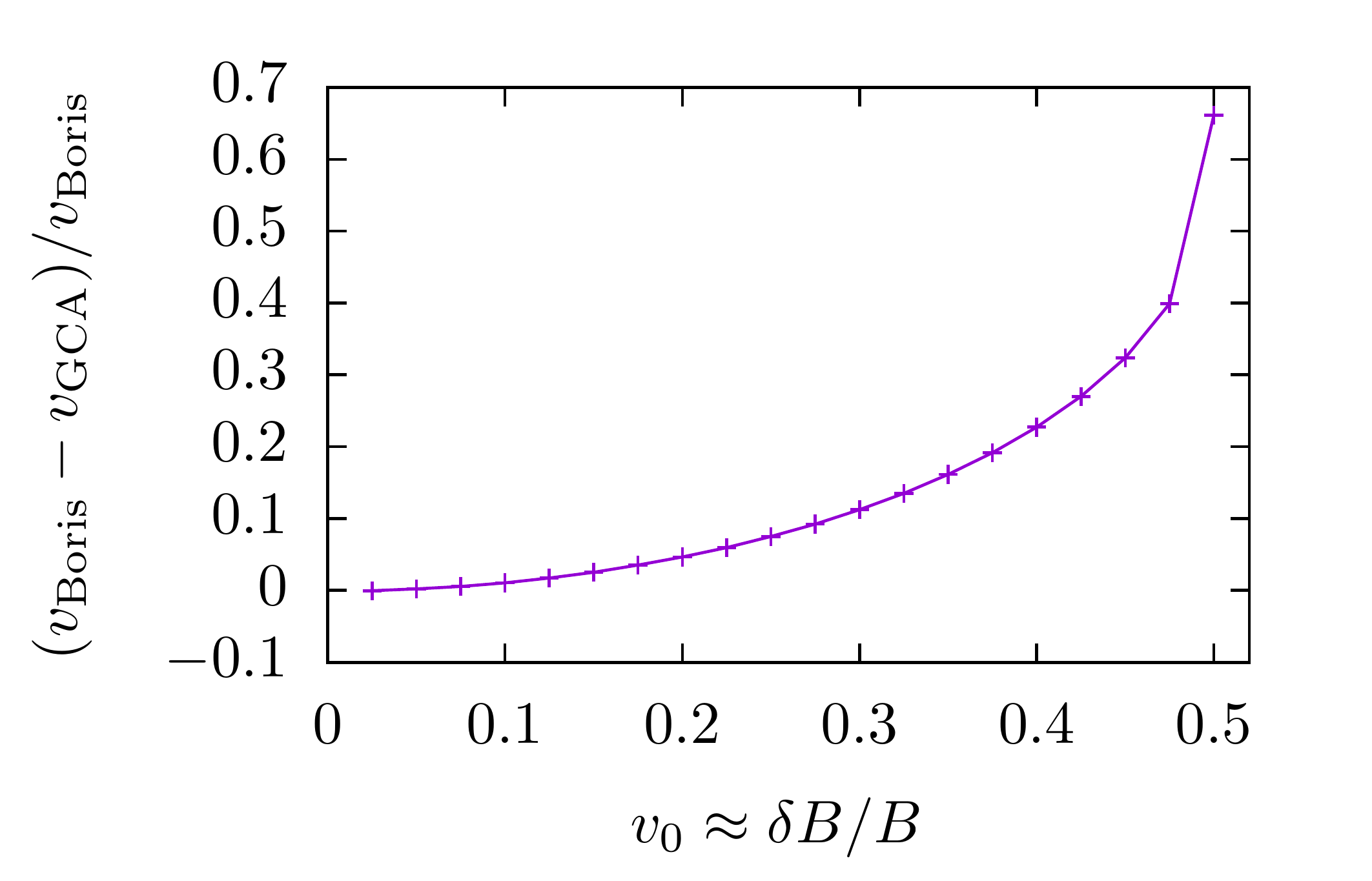}}
  \caption{Left: the $\nabla B$ drift velocity $v_y$ for a magnetic field with a
    linear gradient. Results are shown for Boris method, the guiding center
    approximation and for the approximation of equation \eqref{eq:gradBvel2}.
    Right: The relative difference in the $\nabla B$ drift velocity of the
    GCA compared to Boris method. For this test case
    $\delta B/B \approx v_0$ (omitting SI units).}
  \label{fig:gradB-difference}
\end{figure}

To compare Boris scheme with the GCA, particles are created at the origin, with
an initial velocity $\mathbf{v} = -v_0 \hat{\mathbf{x}}$. Omitting the SI units,
we use $L = 1$, $q/m = 1$ and $B_0 = 1$, so that the gyration radius
$R_c = v_\perp m / (q B) \approx v_0$. By varying $v_0$ the validity of the GCA
changes, since $\delta B/B \approx v_0$. Fig.~\ref{fig:gradB-difference} shows
the difference in the $\nabla B$ drift velocity between GCA and Boris method for
different values of $v_0$. For Boris method $v_y$ was determined by fitting a
line through the local minima of the $y$-coordinate, to ensure samples where
taken at the same gyration-phase. Notice that for a non-relativistic particle
equation \eqref{eq:gradBvel2} and the guiding center approximation give the same
$\nabla B$ drift velocity.

For these tests, a fixed time step of $\Delta t = 5\times 10^{-3}$
was used. The numerical grid contained $128 \times 16 \times 16$ cells, covering
a computational domain of size $2 \, L \times 100 \, L \times 100 \, L$.
The reason for the extra resolution in the $x$-direction is to avoid
interpolation errors in the GCA, which uses extra grid
variables such as $\nabla B$, see section \ref{subsect:GCA}. The linear
interpolation of such terms will not be `exact' when $B_z$ changes sign.

Because of the relatively small time step of $\Delta t = 5\times 10^{-3}$, the
numerical errors in Fig.~\ref{fig:gradB-difference} are negligible compared to
the error due to the guiding center approximation. For $v_0$ up to $0.2$, the
guiding center is still in reasonably good agreement with Boris method, showing
a deviation of less than $5\%$ in the $\nabla B$ drift velocity. However, for
larger $v_0$ (or larger $\delta B/B$) the error increases, and the relative
difference is about $65\%$ for $v_0 = 0.5$.

\subsubsection{Magnetic null}
\label{subsubsect:magnull}

\begin{figure}
  \centering
  \includegraphics{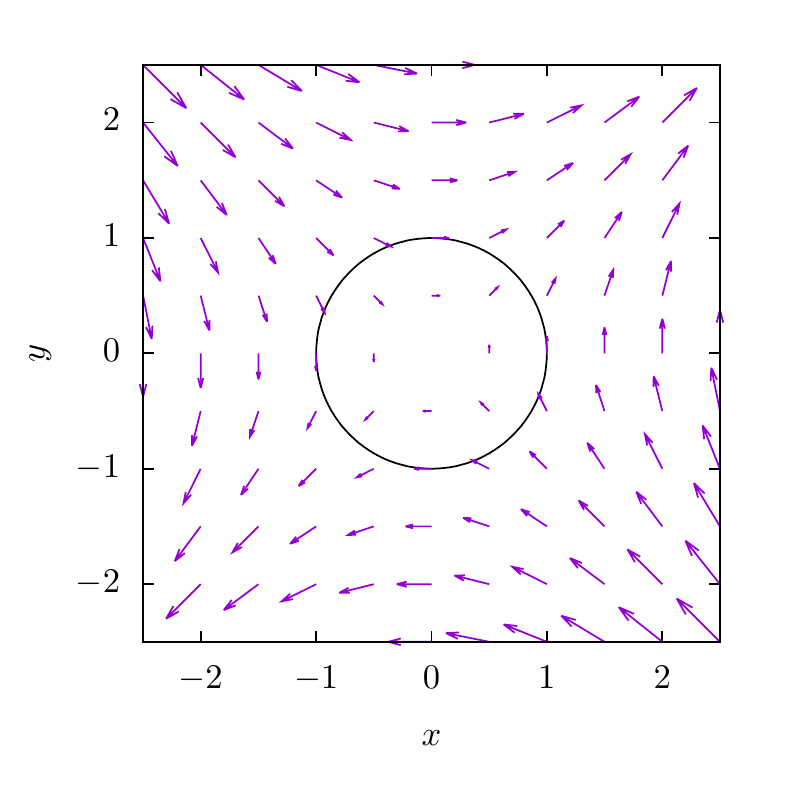}
  \caption{Magnetic field vectors in the $x,y$-plane for the magnetic null test
    case in which $\mathbf{B} = B_0 (y/L, x/L, 0)$. Initially, 500 particles are
    launched from the indicated unit circle, with a radially inwards velocity
    pointing towards the magnetic null.}
  \label{fig:magnull-field}
\end{figure}
\begin{figure}
  \centering
  \includegraphics{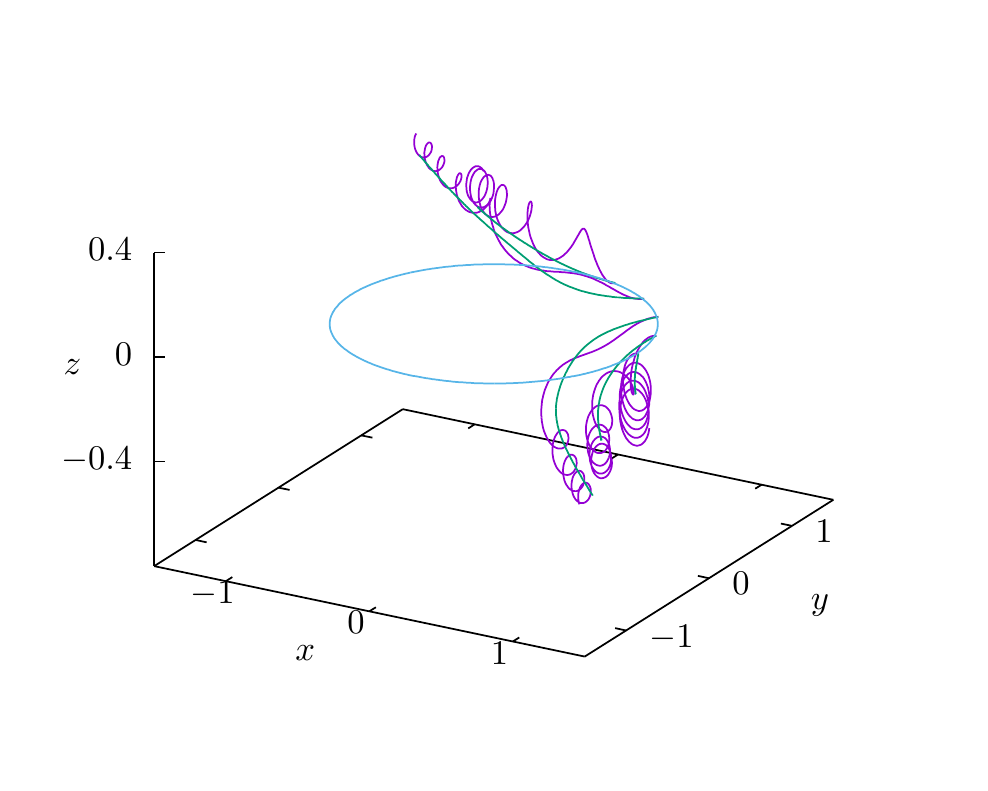}
  \caption{Examples of particle trajectories in the magnetic null case, computed
    with Boris method (purple) and the GCA (green).
    The circle $x^2 + y^2 = 1$ on which particles are initiated is also
    indicated.}
  \label{fig:magnull-example}
\end{figure}

In this example, we consider a magnetic field
\begin{equation}
  \label{eq:magnull-bfield}
  \mathbf{B} = B_0 \, (y/L, x/L, 0),
\end{equation}
where we use (again omitting SI units) $L = 1$, $B_0$ and no electric field
($\mathbf{E} = 0$). The magnetic field, which has a null at the origin, is
illustrated in figure \ref{fig:magnull-field}. Because of the magnetic null, the
GCA is expected to fail when particles get close to the
origin. To investigate this behavior, we place 500 particles on a circle in the
$x,y$-plane, centered around the $z$-axis (i.e., $x^2+y^2=1$ and $z=0$). All
these particles have a purely radial velocity pointing to the origin, of
magnitude $v_r = -0.1$. The particles are then evolved up to
$t = 30$. An example of the resulting trajectories is shown in
Fig.~\ref{fig:magnull-example}, both for the Boris method and the GCA.

In this example, particles are for simplicity created at the same location
regardless of whether GCA or Boris method is used. This leads to an error in
the initial position, since the GCA particles should be initiated at the center
of the gyration. However, the initial error is smallest for particles close to
the diagonals, since their velocity is almost parallel to the magnetic field. We
remark that in many practical applications the magnetic field is not known
beforehand, so precisely matching the guiding centers is difficult.

\begin{figure}
  \centering
  \includegraphics{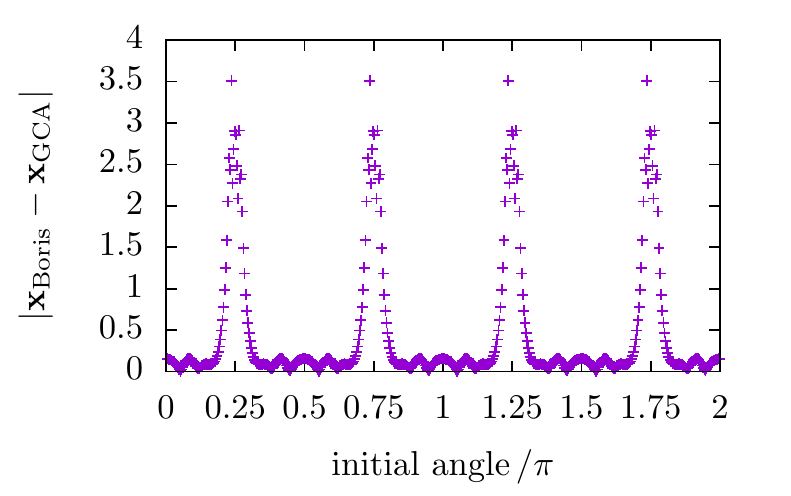}
  \caption{Distance between predicted position at $t = 30$ with
    Boris method and GCA for the magnetic null case. The distance is shown as a
    function of the particles' initial $\phi$-angle on the circle
    $x^2 + y^2 = 1$. Particles close to the diagonals (at
    $\phi \approx \pi/4, 3\pi/4, 5\pi/4, 7\pi/4$) move towards the magnetic
    null, where the GCA breaks down.}
  \label{fig:magnull-difference}
\end{figure}

When particles are located close to one of the four diagonals, their velocity is
almost parallel to the magnetic field. Therefore, they will propagate towards
the origin, where the GCA becomes problematic. This
behavior is quantified in Fig.~\ref{fig:magnull-difference}, which shows the
distance in particle position at $t = 30$ as computed by the GCA
versus Boris method, for a varying initial angle.

For this test case, a fixed time step of $\Delta t = 5\times 10^{-3}$ was used.
The numerical grid contained $64 \times 64 \times 16$ cells, covering a domain
of size $5\times 5\times 5$. Since the magnetic field has linear gradients, it
can be interpolated `exactly' using linear interpolation. However, for some of
the additional grid variables used in GCA method there will be an interpolation
error proportional to $\Delta x^2$. This interpolation error is not the cause
for the difference observed in Fig.~\ref{fig:magnull-difference}, which we have
verified by running a test at a twice higher resolution that produced nearly
identical results.



\subsubsection{Dipolar magnetic field}
\label{subsect:dipole}

The magnetic field surrounding a star or a planet, like Earth, can often be
approximated by a dipole. A pure dipole has no azimuthal component and is
expressed in spherical coordinates by
\begin{equation}
\mathbf{B}(r,\theta) = \frac{M}{r^3} \left[2\cos{\theta}\mathbf{\hat{r}} + \sin\left(\theta\right) \boldsymbol{\hat{\theta}}\right]
\label{eq:dipolefield}
\end{equation}
where $r$ is the radial distance from the center of the dipole, $\theta$ is the
polar angle measured from the dipole axis and $M$ is the dipole moment.
Converting this divergence-free field to Cartesian coordinates gives
\begin{equation}
\mathbf{B}(x,y,z) = \frac{M}{\left(x^2+y^2+z^2\right)^{\frac{5}{2}}} \left[3zx \mathbf{\hat{x}}+ 3zy \mathbf{\hat{y}} + \left(2z^2 -x^2 -y^2\right)\mathbf{\hat{z}}\right].
\label{eq:dipolefield_cart}
\end{equation}

\begin{figure}
  \centering
  \includegraphics{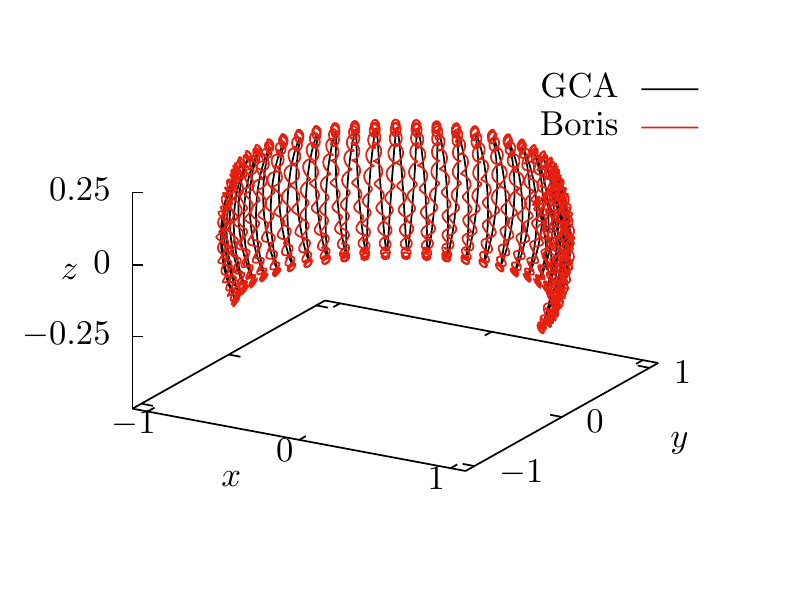}
  \caption{Example of the trajectory of a particle with $q M/m = 40$ in a
    dipolar magnetic field, using Boris method and GCA. The trajectories are
    shown up to $t = 80$.}
  \label{fig:dipole-example}
\end{figure}

Ignoring gyration, we can estimate the gradient-curvature drift of the particle
analytically. The drift velocity results from the third and fifth term in
equation (\ref{eq:gcanewton1}) and can in the absence of volume currents be
written as~\citep{Bittencourt}
\begin{equation}
  \mathbf{v_{R+\nabla B}} = -\frac{m}{qB^4} \left(v^2_{\|} + \frac{1}{2}v^2_{\perp}\right) (\nabla B^2/2) \times \mathbf{B}.
\label{eq:grad_curv_drift}
\end{equation}
For the field of equation \eqref{eq:dipolefield} this leads to a drift motion in
the $\phi$-direction, for which the period is approximately (within
$\sim 0.5\%$) given by~\citep{Walt_1994}:
\begin{equation}
  \label{eq:dipole-period}
  T_\mathrm{dipole} \approx \frac{2 \pi q M}{m v^2 R_0}
  \left[1-\tfrac{1}{3}\sin(\alpha_\mathrm{eq})^{0.62}\right],
\end{equation}
where $R_0$ is the equatorial distance to the guiding center,
$v^2 = v^2_{\|} + v^2_{\perp}$, and
$\alpha_\mathrm{eq} = \tan^{-1}(v_{\perp}/v_{\|})$ the pitch angle at the
equator.
We now test how well the GCA can describe particles in a dipolar field. For
these tests the relative strength of the dipole is varied, which depends on the
ratio $q M/m$. Omitting SI units, we take $q M/m = 10$ up to $100$. Particles
are placed so that their guiding center is located at $\mathbf{x} = (1,0,0)$,
with an initial velocity $\mathbf{v} = (0, 1, 1/2)$. All particles thus have the
same equatorial pitch angle $\alpha_\mathrm{eq}$. An example of the resulting
trajectories is shown in Fig.~\ref{fig:dipole-example}, for both the GCA as
Boris method. Particles exhibit both a mirror motion and a rotation in the
$\phi$-direction. We have used a numerical grid of $256^3$ cells, covering a
computational domain of size $3\times 3 \times 3$, and a fixed time step of
$\Delta t = 5\times 10^{-4}$.

\begin{figure}
  \centering
  \includegraphics{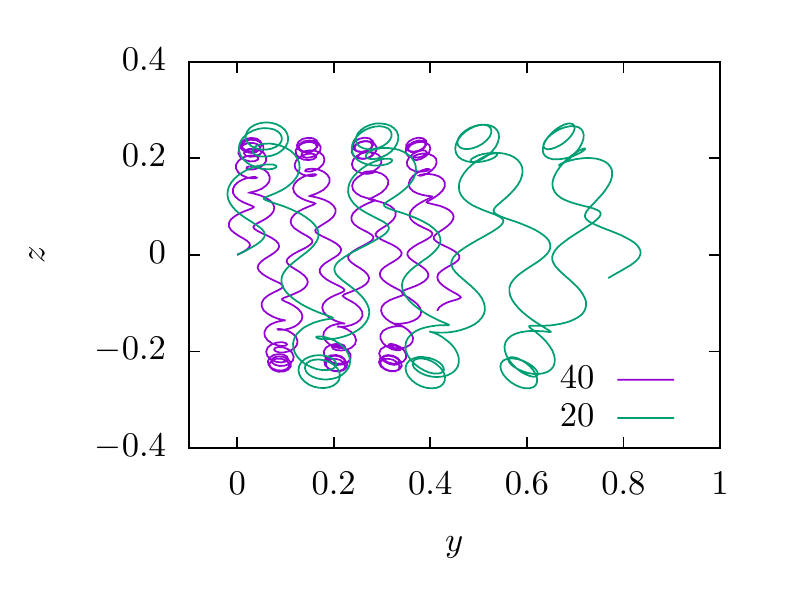}
  \caption{Particle trajectories in a dipolar magnetic field for $q M/m = 20$
    and $40$, computed with Boris method. The trajectories are shown up to
    $t = 10$ and projected along the $x$-axis onto the
    $y,z$-plane.}
  \label{fig:dipole-effect-q}
\end{figure}

\begin{figure}
  \centering
  \subfloat{\includegraphics[width=0.5\columnwidth]{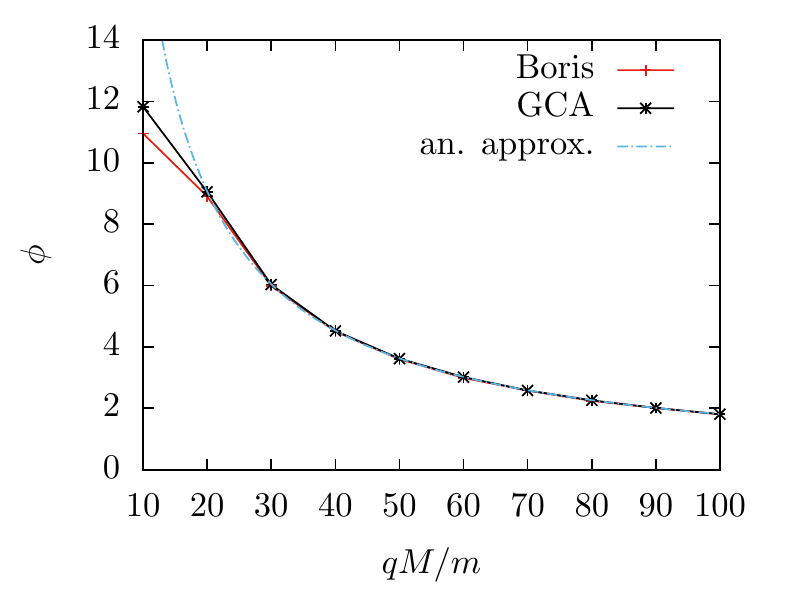}}
  \subfloat{\includegraphics[width=0.5\columnwidth]{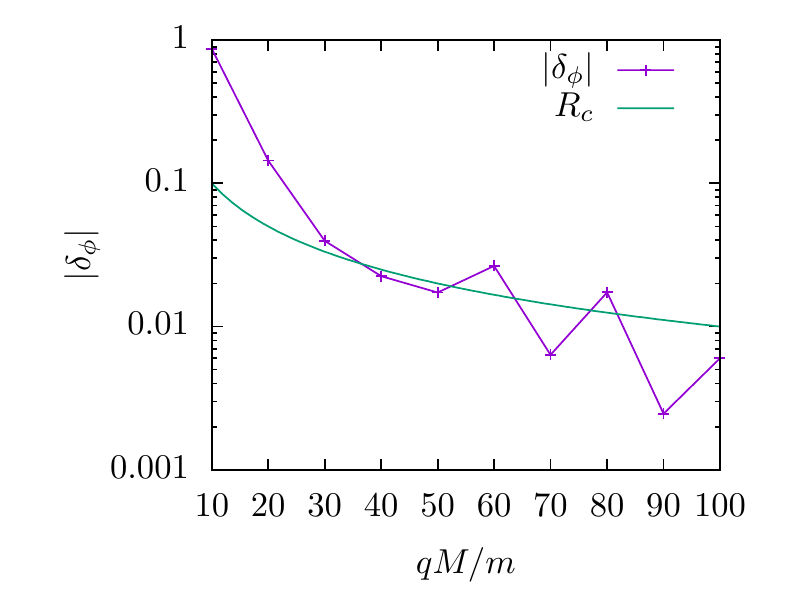}}
  \caption{Left: the $\phi$-angle after $t = 100$ for Boris method and GCA, for
    particles of varying $q M/m$ in a dipolar magnetic field. An analytic
    approximation $\phi = 2\pi \, (100/T_\mathrm{dipole})$ using equation
    \eqref{eq:dipole-period} is also shown. Right: The difference in
    $\phi$-angle at $t = 100$ between Boris method and GCA. The error in
    measuring $\phi$ for Boris method is on the order of the gyroradius $R_c$,
    which is also indicated.}
  \label{fig:dipole-angle}
\end{figure}

Fig.~ \ref{fig:dipole-effect-q} shows the trajectories of particles with
$q M/m = 20$ and $40$ up to $t = 10$, projected on to the $y,z$-plane. A two
times larger value for $q M/m$ leads to half the rotation velocity in the
$\phi$-direction, as also predicted by equation \eqref{eq:dipole-period}. Fig.~
\ref{fig:dipole-angle} shows how the final angle varies with $q M/m$, for Boris
method and GCA, and also shows the result based on equation
\eqref{eq:dipole-period}, namely $\phi = 2\pi \, (100/T_\mathrm{dipole})$.
Because the gyration radius $R_c$ is inversely proportional to $q M/m$, the GCA
should improve for larger $q M/m$. Fig.~\ref{fig:dipole-angle} also shows the
difference in $\phi$-angle after $t = 100$ between GCA and Boris method. The
agreement clearly improves up to $qM/m = 50$, after which the difference
oscillates while still decreasing. The reason for this is that angles were
measured at $t = 100$ as $\phi = \tan^{-1}(y/x)$ (correcting for completed
periods). For Boris method $x,y$ oscillate due to the particle gyration, so that
the error in measuring $\phi$ is approximately the gyroradius $R_c$, which is
also indicated in the figure.

In summary, we find that the GCA approximates the gradient-curvature drift in a
dipolar field to high accuracy for sufficiently large $q M/m$. The relative
error compared to Boris method is around $1.6\%$ for $q M/m = 20$, and rapidly
decreases for larger values of $q M/m$.

\section{Conclusions}
\label{sect:conclusions}
We performed a detailed comparison between several numerical methods to solve for charged particle motion in electromagnetic fields. We compared three explicit leap-frog methods (Boris, Vay and HC), which differ in their choice of the average velocity at half timesteps, with a new implicit solution of the discretized equation of motion. The latter introduces the only average velocity expression which is fully consistent with energy conservation. These four methods to solve the Lorentz equation of motion are further compared to an adaptive Runge-Kutta integration of the relativistic version of the GCA equations. Tests deliberately explore the regime of ultra-relativistic motions, where differences between the obtained numerical solutions become most pronounced.

Tests in uniform fields show that parallel electric field acceleration alone shows only marginal differences between these five approaches, especially in reproducing the exact proportionality between Lorentz factor and time. For particles rapidly accelerating to high Lorentz factors there can be offsets in the computed particle positions for the explicit methods. Ultrarelativistic gyration in a uniform magnetic field demonstrates the conservation of the Lorentz factor (and hence the gyroradius) most convincingly for both the Boris and the implicit scheme. A larger error is found in the steadily increasing phase lag of the gyration, where the HC scheme improves on the Boris, Vay, or implicit strategies.  A test designed to quantify the potential weakness of all schemes for handling the equation of motion analyzes the case of a uniformly moving particle which experiences a net zero Lorentz force. At a Lorentz factor of $\gamma=10^6$, all except the (here trivial) GCA approach show sizeable deviations in position and velocity, with the largest errors when using the Boris algorithm, and the smallest ones when using the implicit scheme. All schemes keep $\gamma$ constant, but introduce a spurious velocity component orthogonal to the initial motion.  A final test in orthogonal uniform electromagnetic fields concentrates on the $\mathbf{E}\times\mathbf{B}$ drift, and at high Lorentz factors only the new implicit method recovers the correct constant gyration radius and Lorentz factor in the comoving frame, over multiple full gyration periods.

Extensions to non-uniform, static magnetic field configurations addressed issues related to magnetic mirroring and gradient-curvature drifts in idealized field prescriptions of astrophysical relevance. In a magnetic mirror (bottle) configuration, a trapped particle can maintain its Lorentz factor to machine precision when using Boris or implicit treatments. While the GCA approximation maintains the magnetic moment by construction, all solution methods for the Lorentz equation show sizeable variations during each cycle through the bottle, and this is most notably influenced by whether analytic or interpolated electromagnetic fields are used.  Field interpolations introduce larger deviations in the magnetic moment, and the Vay scheme in particular performs worst in this aspect. Addressing interpolation effects is particularly relevant for the practical use of these schemes in Particle-in-Cell or MHD codes. The final three tests concentrated on Newtonian regimes, where all Lorentz solvers perform identically, and where we specifically concentrate on the breakdown of the GCA approximation. This was shown to deviate from the expected $\nabla B$ drift velocity in space-dependence magnetic fields, as soon as magnetic fields vary significantly over a gyration period. In such cases, the use of a full Lorentz solver becomes mandatory. The GCA approach is also compared with the Lorentz solver around a magnetic null point, a situation which is of prime importance for particle acceleration in reconnecting fields. This demonstrated that significant errors in the particle positions are obtained through GCA, in particular for particles approaching the magnetic null. Finally, charged particle motions in dipolar fields can be handled well by GCA approximation, and recover the azimuthal drift along with the mirror motion as estimated by theory.

All these methods are implemented in the open source {\tt MPI-AMRVAC} framework (\citealt{Keppensporth}; \citealt{Xia_2017}), and can be used to analyze particle dynamics in evolving electromagnetic fields from MHD simulations. The extension of the methods presented here to general relativistic, covariant formulations is planned for future work in the general relativistic MHD code {\tt BHAC} (\citealt{BHAC}). The implicit particle pusher that is briefly presented here is extended to the fully implicit relativistic Particle-in-Cell code {\tt xPic} (Bacchini et al., in prep).
\acknowledgments
This research was supported by projects GOA/2015-014 (2014-2018 KU Leuven) and the Interuniversity Attraction Poles Programme by the Belgian Science Policy Office (IAP P7/08 CHARM). FB is also supported by the Research Fund KU Leuven and Space Weaves RUN project. JT acknowledges support by postdoctoral fellowship 12Q6117N from Research
Foundation -- Flanders (FWO). CX acknowledges support by postdoctoral fellowship 12C2716N from Research Foundation -- Flanders (FWO). OP is supported by the ERC synergy grant `BlackHoleCam: Imaging the Event Horizon of Black Holes' (Grant No. 610058). LS acknowledges support from DoE DE-SC0016542, NASA Fermi NNX16AR75G, NASA ATP NNX-17AG21G, NSF ACI-1657507, and NSF AST- 1716567. BR would like to thank Anatoly Spitkovsky and Jerome P\'{e}tri for useful comments and suggestions.

\software{{\tt MPI-AMRVAC}, \cite{Keppensporth}, \cite{Xia_2017}, {\tt BHAC}, \cite{BHAC}, {\tt xPic}, Bacchini et al., in prep}.




\bibliographystyle{apj}
\bibliography{mylib3,astro} 


\appendix

\section{Formal proof of energy conservation}
\label{sec:energyconservation}

To formally prove energy conservation for our implicit particle mover we repeat the argument of \cite{Noguchi2007} for the relativistic equation of motion. Starting from the discretized equation of motion where $n$ and $n+1$ indicate consecutive time levels
\begin{equation}
\frac{\mathbf{u}^{n+1} - \mathbf{u}^{n}}{\Delta t} = \frac{q}{m} \left(\mathbf{E}\left(\mathbf{x}^{n+1/2}\right) + \bar{\mathbf{v}} \times \mathbf{B}\left(\mathbf{x}^{n+1/2}\right)\right),
\label{eq:Lorentz_discretized}
\end{equation} 
and taking the dot product with some undefined average velocity $\bar{\mathbf{v}}$ on both sides
\begin{equation}
\bar{\mathbf{v}} \cdot \left(\mathbf{u}^{n+1} - \mathbf{u}^{n}\right) = \frac{q \Delta t}{m} \bar{\mathbf{v}} \cdot {\mathbf{E}}\left(\mathbf{x}^{n+1/2}\right).
\label{eq:Lorentzdotv}
\end{equation}
The magnetic field does not exert work on a particle and the work done by an electric field is 
\begin{equation*}
W_E = q \mathbf{E}\left(\mathbf{x}^{n+1/2}\right) \cdot \ \bar{\mathbf{v}} \Delta t = 
\end{equation*}
\begin{equation*}
q \mathbf{E}\left(\mathbf{x}^{n+1/2}\right) \cdot \left(\mathbf{x}^{n+1} - \mathbf{x}^{n}\right) =
\end{equation*}
\begin{equation}
m c^2 \left( \gamma^{n+1} - \gamma^n\right),
\label{eq:energyderivation}
\end{equation}
where we use the definition of work as the difference in kinetic energy $W_E = m c^2 \left(\gamma^{n+1} - 1\right) - m c^2 \left(\gamma^n - 1\right)$. This reduces to
\begin{equation}
\bar{\mathbf{v}} \cdot \left(\mathbf{u}^{n+1} - \mathbf{u}^{n}\right) = (\gamma^{n+1} - \gamma^n)c^2,
\label{eq:conditionvbar}
\end{equation}
and gives us an energy argument to determine how $\bar{\mathbf{v}}$ has to be chosen to obey energy conservation for the particle mover. 

\subsection{Implicit midpoint scheme}
Plugging in the velocity at half timestep as used by the fully implicit scheme
\begin{equation}
\bar{\mathbf{v}} = \frac{\mathbf{u}^{n+1} + \mathbf{u}^{n}}{\gamma^{n+1} + \gamma^n}
\label{eq:vsolution}
\end{equation}
into equation (\ref{eq:conditionvbar}) gives
\begin{equation}
\frac{(u^{n+1})^2 - (u^n)^2}{\gamma^{n+1} + \gamma^n} = (\gamma^{n+1} - \gamma^n)c^2,
\end{equation}
and by using the definition of $\gamma$ in terms of momentum for discretized Lorentz factors $(\gamma^{n+1})^2 = (u^{n+1})^2/c^2 + 1$ and $(\gamma^{n})^2 = (u^{n})^2/c^2 + 1$ we prove that equality (\ref{eq:conditionvbar}) is satisfied for (\ref{eq:vsolution}).

\subsection{Boris scheme}
The average velocity for the Boris scheme is given by
\begin{equation}
\bar{\mathbf{v}} = \frac{\mathbf{u}^{n+1} + \mathbf{u}^n}{2 \gamma^{n+1/2}}
\end{equation}
with 
\begin{equation}
\gamma^{n+1/2} = \sqrt{1+(u^{n+1/2})^2/c^2} = \sqrt{1+(u^{-})^2/c^2}
\end{equation} 
Plugging this $\bar{\mathbf{v}}$ into equation (\ref{eq:conditionvbar}) we obtain
\begin{equation*} 
\frac{\mathbf{u}^{n+1} + \mathbf{u}^n}{2 \gamma^{n+1/2}} \cdot \left(\mathbf{u}^{n+1} - \mathbf{u}^{n}\right)  = 
\end{equation*}
\begin{equation}
\frac{(u^{n+1})^2 - (u^n)^2}{2\gamma^{n+1/2}} = (\gamma^{n+1} - \gamma^n)c^2,
\end{equation}
and by following the same procedure as for the implicit scheme we find
\begin{equation*}
\frac{\left(\gamma^{n+1}\right)^2 - \left(\gamma^n\right)^2}{2\gamma^{n+1/2}} = 
\end{equation*}
\begin{equation}
\frac{\left(\gamma^{n+1} + \gamma^n\right)\left(\gamma^{n+1} - \gamma^n\right)}{2\gamma^{n+1/2} } = \gamma^{n+1} - \gamma^n.
\end{equation}
This equation only holds in the specific case of $(\gamma^{n+1} + \gamma^n)/2 = \gamma^{n+1/2}$. Using the definition of $(\gamma^{n+1/2})^2 = 1 + \left(\mathbf{u}^{n} + \boldsymbol{\epsilon}\right)^2/c^2 = 1 + \left(\mathbf{u}^{n+1} - \boldsymbol{\epsilon}\right)^2/c^2$, with $\boldsymbol{\epsilon} = (q \Delta t/2m)\mathbf{E}(\mathbf{x}^{n+1/2})$ one can show that
\begin{equation}
\frac{(\mathbf{u}^{n+1})^2 - (\mathbf{u}^n)^2}{\sqrt{1+(\mathbf{u}^{n} + \boldsymbol{\epsilon})^2/c^2} + \sqrt{1+(\mathbf{u}^{n+1} - \boldsymbol{\epsilon})^2/c^2}} = \left(\gamma^{n+1} - \gamma^n\right)c^2,
\end{equation}
which is only true in the case
\begin{equation*}
\sqrt{1+\left(\left(\mathbf{u}^n\right)^2 + 2\mathbf{u}^n \cdot \boldsymbol{\epsilon} + \epsilon^2\right)/c^2} + \sqrt{1+\left(\left(\mathbf{u}^{n+1}\right)^2 - 2\mathbf{u}^{n+1} \cdot \boldsymbol{\epsilon} + \epsilon^2\right)/c^2} = 
\end{equation*}
\begin{equation}
\sqrt{(\gamma^n)^2 + \left(2\mathbf{u}^n \cdot \boldsymbol{\epsilon} + \epsilon^2\right)/c^2} + \sqrt{(\gamma^{n+1})^2 + \left(\epsilon^2 - 2\mathbf{u}^{n+1} \cdot \boldsymbol{\epsilon}\right)/c^2} =  \gamma^{n+1} + \gamma^n.
\end{equation}
The equality is only satisfied in specific cases, e.g. the case of no electric field $\boldsymbol{\epsilon}  = \mathbf{0}$, which is trivial since a magnetic field does not exert work on a particle. The Boris scheme is therefore energy conserving for a vanishing electric field. However that does not mean that a high accuracy of energy conservation cannot be obtained with a nonzero electric field.

\subsection{Vay scheme}
For the Vay scheme the choice of the average velocity is given by $\bar{\mathbf{v}} = (\mathbf{v}^{n+1} + \mathbf{v}^{n})/2 = (\mathbf{u}^{n+1}/\gamma^{n+1} + \mathbf{u}^{n}/\gamma^{n})/2 $ (\citealt{Vay_2008}). Plugging this $\bar{\mathbf{v}}$ into equation (\ref{eq:conditionvbar}) we obtain
\begin{equation*}
\left(\frac{\mathbf{u}^{n+1}}{\gamma^{n+1}} + \frac{\mathbf{u}^{n}}{\gamma^n}\right) \cdot \left(\mathbf{u}^{n+1} - \mathbf{u}^{n}\right) /2 = 
\end{equation*}
\begin{equation*}
\frac{1}{2} \left[\frac{\left(u^{n+1}\right)^2}{\gamma^{n+1}} - \frac{\left(u^n\right)^2}{\gamma^n} + \frac{\mathbf{u}^{n+1} \cdot \mathbf{u}^{n}}{\gamma^{n}} - \frac{\mathbf{u}^{n+1} \cdot \mathbf{u}^{n}}{\gamma^{n+1}}\right] = 
\end{equation*}
\begin{equation*}
\frac{1}{2}\left[\gamma^{n+1} c^2-\gamma^n c^2 - \frac{c^2}{\gamma^{n+1}} + \frac{c^2}{\gamma^n} + \mathbf{u}^n \cdot \mathbf{u}^{n+1}\left(\frac{1}{\gamma^n} - \frac{1}{\gamma^{n+1}} \right)\right] = 
\end{equation*}
\begin{equation}
\left(\gamma^{n+1} - \gamma^n\right) \left(\frac{c^2}{2}+\frac{1}{2}\frac{\mathbf{u}^n\cdot\mathbf{u}^{n+1}+c^2}{\gamma^n \gamma^{n+1}}\right) \leq (\gamma^{n+1} - \gamma^n)c^2.
\end{equation}
The equality is only true in the very specific case where $(\mathbf{u}^n \cdot \mathbf{u}^{n+1} + c^2)/(\gamma^n \gamma^{n+1}) = c^2$. The equality is satisfied if $\mathbf{u}^n = \mathbf{u}^{n+1}$, which is the trivial case where the particles energy and momentum do not change. When energy conservation is not satisfied to machine precision, as is generally the case for this choice of $\bar{\mathbf{v}}$ the particles are spuriously heated. In practice the scheme compute particle dynamics very accurately, with bounded energy errors, but energy is not conserved in the strict sense. In an implicit scheme, based on the Vay framework (\citealt{Petri2017}), the choice of the timestep will not change this, however the number of iterations in the implicit step can result in a high accuracy for energy conservation.

\subsection{Higuera-Cary scheme}

In the Higuera-Cary scheme another average velocity is derived, that is proven to result in a volume preserving method (\citealt{HC_2017})
\begin{equation}
\bar{\mathbf{v}} = \frac{\mathbf{u}^{n+1} + \mathbf{u}^n}{2 \bar{\gamma}}
\end{equation}
\begin{equation}
\bar{\gamma} = \sqrt{1 +\left(\frac{\mathbf{u}^{n+1} + \mathbf{u}^n}{2c}\right)^2}.
\end{equation} 
Plugging this average velocity into equation (\ref{eq:conditionvbar}) results in the same final condition as for the choice of the average velocity in the Boris scheme
\begin{equation}
\frac{\left(\gamma^{n+1} + \gamma^n\right)\left(\gamma^{n+1} - \gamma^n\right)}{2\gamma^{n+1/2} } = \gamma^{n+1} - \gamma^n.
\end{equation}
This is only satisfied if
\begin{equation}
2\sqrt{1+\left(\frac{\mathbf{u}^{n+1}+\mathbf{u}^n}{2c}\right)^2} = \gamma^{n+1} + \gamma^{n},
\end{equation}
however
\begin{equation*}
2\sqrt{1+\left(\frac{\mathbf{u}^{n+1}+\mathbf{u}^n}{2c}\right)^2} = \sqrt{\left[1+(\mathbf{u}^{n+1})^2/c^2\right] + \left[1+(\mathbf{u}^{n})^2/c^2\right] + 2\left(1+\mathbf{u}^{n+1} \cdot \mathbf{u}^n/c^2\right)} = 
\end{equation*}
\begin{equation}
\sqrt{(\gamma^{n+1})^2 + (\gamma^{n})^2 + 2\left(1+\left(\mathbf{u}^{n+1} \cdot \mathbf{u}^n\right)/c^2\right)} \leq \gamma^{n+1} + \gamma^n,
\end{equation}
where the equality is only satisfied in the trivial case of a non-varying particle momentum $\mathbf{u}^n = \mathbf{u}^{n+1}$, resulting in the same condition as for the Vay scheme.




\label{lastpage}
\end{document}